\documentclass[]{article}

\voffset= - 1.0in
\hoffset= - 1.0in         % switch off for draft style

\usepackage{amssymb}

\font\numbers=cmss12
\font\upright=cmu10 scaled\magstep1
\def\stroke{\vrule height8pt width0.4pt depth-0.1pt}
\def\topfleck{\vrule height8pt width0.5pt depth-5.9pt}
\def\botfleck{\vrule height2pt width0.5pt depth0.1pt}
\def\Zmath{\vcenter{\hbox{\numbers\rlap{\rlap{Z}\kern 0.8pt\topfleck}\kern
2.2pt
                   \rlap Z\kern 6pt\botfleck\kern 1pt}}}
\def\Qmath{\vcenter{\hbox{\upright\rlap{\rlap{Q}\kern
                   3.8pt\stroke}\phantom{Q}}}}
\def\Nmath{\vcenter{\hbox{\upright\rlap{I}\kern 1.7pt N}}}
\def\Cmath{\vcenter{\hbox{\upright\rlap{\rlap{C}\kern
                   3.8pt\stroke}\phantom{C}}}}
\def\Rmath{\vcenter{\hbox{\upright\rlap{I}\kern 1.7pt R}}}
\def\Z{\ifmmode\Zmath\else$\Zmath$\fi}
\def\Q{\ifmmode\Qmath\else$\Qmath$\fi}
\def\N{\ifmmode\Nmath\else$\Nmath$\fi}
\def\C{\ifmmode\Cmath\else$\Cmath$\fi}
\def\R{\ifmmode\Rmath\else$\Rmath$\fi}
\def\d{\partial}

\def\bea{\begin{eqnarray}}
\def\eea{\end{eqnarray}}

\def\beq{\begin{equation}}
\def\eeq{\end{equation}}
\def\ba{\beq\begin{array}{c}}
\def\ea{\end{array}\eeq}
\def\be{\ba}
\def\ee{\ea}
\def\stackreb#1#2{\mathrel{\mathop{#2}\limits_{#1}}}
\def\Tr{{\rm Tr}}

\def\Bf#1{\mbox{\boldmath $#1$}}
\def\balpha{{\Bf\alpha}}

\def\bgamma{{\Bf\gamma}}

\def\bPhi{{\Bf\Phi}}

\def\bxi{{\Bf\xi}}

\def\2{{1\over 2}}
\def\N2{${\cal N}=2$}
\def\4N{${\cal N}=4$}
\def\1N{${\cal N}=1$}

\makeatletter
\newdimen\normalarrayskip              % skip between lines
\newdimen\minarrayskip                 % minimal skip between lines
\normalarrayskip\baselineskip
\minarrayskip\jot
\newif\ifold             \oldtrue            \def\new{\oldfalse}
\def\arraymode{\ifold\relax\else\displaystyle\fi} % mode of array entries
\def\eqnumphantom{\phantom{(\theequation)}}     % right phantom in eqnarray
\def\@arrayskip{\ifold\baselineskip\z@\lineskip\z@
     \else
     \baselineskip\minarrayskip\lineskip2\minarrayskip\fi}
\def\@arrayclassz{\ifcase \@lastchclass \@acolampacol \or
\@ampacol \or \or \or \@addamp \or
   \@acolampacol \or \@firstampfalse \@acol \fi
\edef\@preamble{\@preamble
  \ifcase \@chnum
     \hfil$\relax\arraymode\@sharp$\hfil
     \or $\relax\arraymode\@sharp$\hfil
     \or \hfil$\relax\arraymode\@sharp$\fi}}
\def\@array[#1]#2{\setbox\@arstrutbox=\hbox{\vrule
     height\arraystretch \ht\strutbox
     depth\arraystretch \dp\strutbox
     width\z@}\@mkpream{#2}\edef\@preamble{\halign
\noexpand\@halignto
\bgroup \tabskip\z@ \@arstrut \@preamble \tabskip\z@ \cr}%
\let\@startpbox\@@startpbox \let\@endpbox\@@endpbox
  \if #1t\vtop \else \if#1b\vbox \else \vcenter \fi\fi
  \bgroup \let\par\relax
  \let\@sharp##\let\protect\relax
  \@arrayskip\@preamble}
\def\eqnarray{\stepcounter{equation}%
              \let\@currentlabel=\theequation
              \global\@eqnswtrue
              \global\@eqcnt\z@
              \tabskip\@centering
              \let\\=\@eqncr
 \halign to \displaywidth\bgroup
    \eqnumphantom\@eqnsel\hskip\@centering
    $\displaystyle \tabskip\z@ {##}$%
    \global\@eqcnt\@ne \hskip 2\arraycolsep
         $\displaystyle\arraymode{##}$\hfil
    \global\@eqcnt\tw@ \hskip 2\arraycolsep
         $\displaystyle\tabskip\z@{##}$\hfil
         \tabskip\@centering
    &{##}\tabskip\z@\cr}
\begingroup\ifx\undefined\newsymbol \else\def\input#1 {\endgroup}\fi

\def\ba{\beq\new\begin{array}{c}}
\def\ea{\end{array}\eeq}
\def\be{\ba}
\def\ee{\ea}
\def\stackreb#1#2{\mathrel{\mathop{#2}\limits_{#1}}}
\def\Tr{{\rm Tr}}

\begin{document}

\begin{flushright}
FIAN/TD-5/2000\\
ITEP/TH-25/2000\\
MS-00-009\\
hep-th/0009060
\end{flushright}
\vspace{0.5cm}
\begin{center}
{\LARGE \bf Singular Phases of Seiberg-Witten Integrable Systems:
Weak and Strong Coupling}\\
\vspace{0.5cm}
{\Large H.W.Braden\footnote{Department of Mathematics and Statistics,
University of Edinburgh, Edinburgh EH9 3JZ Scotland;
e-mail address: hwb@ed.ac.uk }} {\large and} {\Large
A.Marshakov\footnote{Theory
Department, Lebedev Physics Institute, Moscow
~117924, Russia and ITEP,
Moscow 117259, Russia; e-mail address: mars@lpi.ru;
andrei@heron.itep.ru}}
\\
\end{center}

\begin{quotation}
We consider the singular phases of the smooth finite-gap integrable systems
arising in the context of Seiberg-Witten theory.
These degenerate limits correspond to the weak and strong coupling
regimes of SUSY gauge theories. The spectral curves in such limits
acquire simpler forms: in most cases they become rational, and
the corresponding expressions for coupling constants and superpotentials
can be computed explicitly. We verify that in accordance with the computations
from quantum field theory, the weak-coupling limit
gives rise to precisely the ``trigonometric" family of
Calogero-Moser and open Toda models, while the strong-coupling limit
corresponds to the solitonic degenerations of the finite-gap
solutions. The formulae arising provide some new insights into the corresponding
phenomena in SUSY gauge theories. Some open conjectures have been proven.
\end{quotation}

\section{Introduction}

Recent progress in understanding non-perturbative structures in
supersymmetric (SUSY) gauge theories \cite{SW,SW+} has also shed new insight
upon the role of integrable structures in modern theoretical physics.
Surprisingly, several integrable systems that were introduced and studied over
the years as simplified models for quantum-mechanical and low-dimensional
field theoretical problems have now reappeared as an important
tool for understanding physically interesting effects in almost realistic,
effective gauge theories.
Part of the reason for this (familiar from supergravity investigations)
is that the scalars of say a $D=4$ \N2 SUSY vector supermultiplet are 
constrained to lie on a special K\"ahler manifold.
With appropriate integrality constraints the cotangent bundle of such
a manifold appears as the phase space of an algebraically completely
integrable system (see, for example \cite{Freed}).
Yet an {\it a priori} argument for which integrable
systems should arise, and what lies at the heart of this
relationship between Seiberg-Witten (SW) theories \cite{SW,SW+}
and integrable systems \cite{GKMMM} 
(see also \cite{don, SWCal} and the 
books \cite{SWIS,SWWE} and references therein)
still remains unanswered, and
presents more questions than answers.
Nonetheless this correspondence is accepted as providing a strong
technical tool for the classification and computation of instanton effects
in \N2 SUSY gauge theories \cite{instcomp,ema}, with integrable systems
yielding insights not yet understood in conventional gauge theoretic
terms.
Indeed one can straightforwardly present a {\em dictionary} for the
correspondence between integrable systems and SW theories, but a majority of
the ``arrows" of the correspondence require further comment, not all of
which is satisfactory.

In what follows we shall study various singular limits of this correspondence.
This will yield both a better physical insight and simpler expressions for
the corresponding phenomena in the quantum gauge theories.
The integrable systems arising in the context of SW theory are
the well-known finite-gap solutions (see, for example \cite{DKN})
associated to complex curves or Riemann surfaces of finite genus.
Though comparatively simple complex manifolds, explicit formulae and
constructions for these integrable systems and their solutions are often
lacking.
However, in certain {\em singular} limits, their structure simplifies
drastically: this happens when one comes to the boundaries of the moduli space
of their complex structures, or, in different terms, to the ``boundary" values
of vacua parameters of the gauge theory. These are the limits we shall
study in this paper. The simplifications arising at these limits
will enable us to derive explicit formulae and constructions, as well as
provide certain insight into the corresponding phenomena in  the
quantum gauge theory.

The boundaries of the moduli spaces of SW curves are characterised by either
the vanishing or divergence of their period matrices $T_{ij}$, which
play the role of couplings in the effective gauge theories.
The limits $T_{ij}\to\infty$ and $T_{ij}\to 0$ correspond to weak and
strong coupling phases of the gauge theory,
and they are related by the modular transformation $T\leftrightarrow T^{-1}$
playing the role of S-duality \cite{SW}.
Below we shall study the corresponding limits. Both limits
appear rather naturally from the perspective of the integrable models.
The weak-coupling limit of the gauge theory, when
instanton contributions are exponentially suppressed and can be neglected,
is almost trivial from the field theory perspective.
The corresponding limit on the integrable systems side is a ``simplification"
of the interparticle potential: the non-periodic (or open) Toda chain
is a limit of the periodic Toda ``molecule"; the
trigonometric Calogero-Moser-Sutherland model is a
degeneration of the elliptic Calogero-Moser model, and so on.
In these cases (and now in fact {\em only} in these cases) can the
SW integral formulae be practically {\em derived} from
perturbative computations in the \N2 gauge theories.

The strong-coupling limit of SW theories corresponds to another limit of the
integrable models, that of smooth finite-gap solutions degenerating into
solitons \cite{AMMinn}. Physically this is the weakly-coupled limit of a dual
theory in which one expects monopole condensation, breaking symmetry down
to \1N, and confinement \cite{SW}.
As we shall see this is in precise correspondence with the
properties of the solitonic solutions of the SW integrable systems, which
correspond to particular values of the integrals of motions. The corresponding
vacuum expectation values (VEV's) are restricted to certain points in the
moduli space,  which can be treated as extrema of an \1N superpotential.

The paper is organised as follows. In section 2 we
begin with a discussion of the
weak-coupling regime in \N2 SUSY gauge theories, presenting some
intuitive motivation for why this can be directly related to the integrable
systems of the Toda or Calogero-Moser families.
Here we also review general properties of
the corresponding finite-gap solutions. In section 3 we discuss in detail the
perturbative limit, giving rise to the {\em open} Toda chain family.
We present explicit formulas for the generating functions for the
open Toda chain (Toda molecule) obtained along these line and discuss
their relations with matrix models and duality in integrable systems.
We discuss in a similar way the perturbative limit of elliptic
Calogero-Moser models. In section 4 we turn to the solitonic or strong-coupling
limit of the finite-gap integrable systems and derive the explicit form
of solution in terms of the Baker-Akhiezer functions. This is
the most physically attractive phase of the SW theories, corresponding
to the confinement in \1N gauge theories, and we show that the phase of
the solitons is related to the string tensions in the confining phase.
We also prove (in Appendix B) a conjecture of Edelstein and Mas  related to this
strong coupling regime. We conclude with a discussion.

\section{\N2 SUSY gauge theories and integrable systems } %\cite{GKMMM}

We begin in this section by reviewing various aspects of the correspondence
between \N2 SUSY gauge theories and integrable systems that we shall need in
our later discussion.
Throughout we will limit our attention to those models arising
in the context of either $D=4$ pure \N2 $SU(N)$ gauge theories, or those
with adjoint matter. These are listed in Figure 1.
Extensions to other gauge groups and matter multiplets
are possible, though will not be our focus here. Indeed, for concreteness
we will describe the $N=2$ cases of these models in some detail.

\begin{center}
\begin{tabular}{lccc}
$d=5$&{ Ruijsenaars-Schneider}&$\longrightarrow$&{`Relativistic' Toda}\\
\\
&$\downarrow$&&$\downarrow$\\
\\
$d=4$&{ Elliptic Calogero-Moser}&$\longrightarrow$&{\rm Periodic Toda}\\
   &{ \N2  SYM with adjoint matter}&&{ Pure \N2 SYM}\\
\\
&$\downarrow$&{\it Perturbative Limits}&$\downarrow$\\
\\
&{ Calogero-Moser-Sutherland}&&{ Open  Toda}\\
\\
\end{tabular}

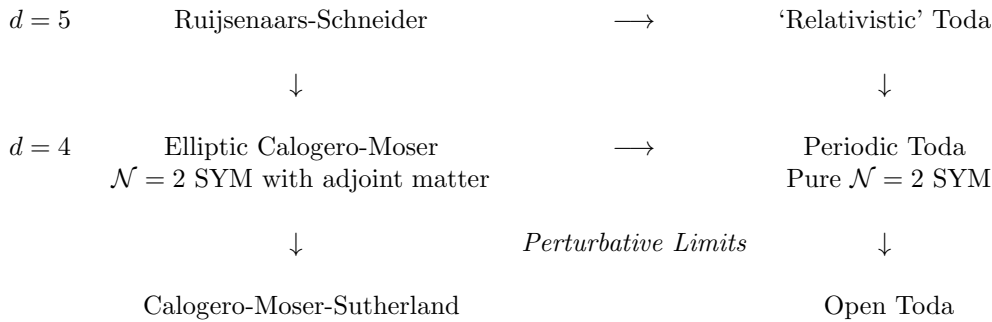
Figure 1.
\end{center}

We will begin the section with some intuitive motivation as to why the
weak-coupling regime in \N2 SUSY gauge theories can be directly
related to the integrable systems of the Toda or Calogero-Moser families.
Then we review general properties of
the corresponding finite-gap solutions and discuss in detail their
perturbative limit, giving rise to the {\em open} Toda chain family.

Although much of what we say in this section is by way of review
we will also present several
new technical details concerning the relation between the Calogero-Moser
and Ruijsenaars-Schneider  spectral curves.
The books \cite{SWIS, SWWE} contain both recent reviews and general
references for aspects of this correspondence not touched upon here.

\subsection{Motivations: Perturbative Calculations and $N=2$ Examples}
%\subsection{Motivations: the perturbative calculations}

The connections between SW theories and integrable systems can be
discussed even at the {\em perturbative} level, where \N2 SUSY
effective actions are completely determined by the 1-loop contributions.
In the most well-known example, that of pure \N2 gauge theory with $SU(N)$
gauge group, the scalar field $\bPhi = \|\Phi_{ij}\|$
of the \N2 vector supermultiplet may acquire nonzero VEV's
$\bPhi = {\rm diag}(a_1,\dots,a_N)$ in extremising the potential
$\Tr [\bPhi,\bPhi^\dagger ]^2 $.
This (generically) breaks the $SU(N)$ gauge group down to the maximal compact
$U(1)^{N-1}$. The masses of $W$-bosons and their superpartners
are (classically) given by $m^W_{ij}= a_{ij}\equiv a_i- a_j$ because
of the interaction term
$\left([A_\mu,\bPhi]_{ij}\right)^2 = \left(A_\mu^{ij}( a_i- a_j)\right)^2$.
By a simple technical trick these masses can be uniformly expressed in terms
of the generating polynomial
\be
w = P_N(\lambda ) = \det_{N\times N} (\lambda - \bPhi) =
\prod_{i=1}^N (\lambda -  a_i),
\label{polyn}
\ee
via the {\em contour integral}
\be\label{pertsample}
m^W_{ij} = \oint_{C_{ij}}dS^{\rm pert} = \oint_{C_{ij}} \lambda d\log w=
\oint_{C_{ij}} \lambda d\log P_{N}(\lambda ).
\ee
For a particular ``figure-of-eight" like contour $C_{ij}$ around the roots
$\lambda =  a_i$ and $\lambda =  a_j$ this may be computed via the residue
formula. This means that the contour integrals in the
complex $\lambda $-plane (\ref{pertsample}) for such
contours around these singular points (the roots of the polynomial
(\ref{polyn}))
give  one set of the BPS masses in the SW theory. Another
set of masses (monopoles) may be associated to {\em dual} contours starting and
ending at the points $a_i$ where $dS^{\rm pert}$ has singularities.
Although such integrals are divergent, the
result of such an integration says that the monopole masses are proportional
to the masses $m^W_{ij}$, multiplied by the inverse square of the coupling.
The divergences are absorbed by a renormalisation of the coupling.
The integrals over the contours on the marked plane we are dealing with here 
are, as we shall later see, to be viewed in terms of contour integrals
on a {\em degenerate} Riemann surface.

Now in \N2 perturbation theory the effective action (or prepotential)
${\cal F}$ and the set of effective couplings $T_{jk}$, related by
$-i\pi T_{jk} = {\d^2{\cal F}\over\d a_j\d a_k}$,
are determined by the 1-loop diagrams. This results in the logarithmic term
\be\label{effcharge}
T_{jk}^{\rm pert} = {1\over 2\pi i}\sum_{({\rm masses M})_{jk}}
 \log {({\rm mass})^2\over\Lambda ^2} =
{1\over 2\pi i}\log {a_{jk}^2\over\Lambda ^2},
\ee
where the scale parameter $\Lambda\equiv\Lambda_{QCD}$ so introduced
may be related to the bare coupling $\tau $. In the perturbative {\em
weak-coupling} limit of the SW construction this is all one has. Instanton
contributions to the prepotential which are proportional to powers of
 $\Lambda^{2N}$ are suppressed, and one keeps only the terms proportional to
$\log\Lambda$ or the bare coupling $\tau $.

It remains to connect this discussion with integrable systems.
The connection comes by interpreting the differential $dS^{\rm pert}$ with
the generating differential of a Hamiltonian system. Remarkably, this
interpretation extends beyond the perturbative regime!
Lets consider the simplest SW theory \cite{SW}, the $SU(2)$ pure gauge theory,
where eq.~(\ref{polyn}) turns into
\be
w = \lambda^2 - h.
\label{polsu2}
\ee
Here $h = \2\Tr\bPhi^2= \2a^2$ and the masses (\ref{pertsample}) are now defined
by the contour integrals of
\be
dS = \lambda d\log w = {\lambda d\lambda\over\lambda - \sqrt{h}} +
{\lambda d\lambda\over\lambda + \sqrt{h}}.
\label{dSop}
\ee
Now,  eqs.~(\ref{polsu2}) and (\ref{dSop}) can be interpreted
as an {\em integration} of a simple dynamical system, the $SL(2)$
open Toda chain (or Liouville model) with Hamiltonian
\be
\label{CM2pert}
H=\frac{1}{2}(p\sp2+e\sp{2q}).
\ee
This has solution $e^q=a/\cosh(a[t-t_0])$, where $a=\sqrt{2 H}$.
One verifies that
\be \frac{p\sp2 dp}{p\sp2-a\sp2}=pdq+\frac{ap da}{p\sp2-a\sp2}
\ee
and so upon integrating over a trajectory
$\oint_\gamma \frac{p\sp2 dp}{p\sp2-a\sp2}=\oint_\gamma pdq=2\pi i a$.
But with the co-ordinate $w=e^{2q}$, momentum $p=\lambda$ and Hamiltonian
(energy) $H=h$ this is essentially (\ref{dSop}). That is,
integration of the canonical
differential $dS =2 pdq$ over the trajectories of the Toda chain solutions gives
rise, for various (complexified) trajectories, to the BPS masses in the
SW theory.

This is actually a general rule. The perturbative mass spectrum and
effective couplings of the \N2 theories of the ``SW family" appear as actions
when integrating the canonical differential of
``open" or trigonometric families of integrable systems,
the open Toda chain, the trigonometric Calogero-Moser and the trigonometric
Ruijsenaars-Schneider systems.
For pure gauge theories we have the $N$-particle (open) Toda chain with
(rational) curves  given by eq.~(\ref{polyn}) and generating differential given
by (\ref{pertsample}). The inclusion of adjoint matter is associated to the
trigonometric Calogero-Moser-Sutherland model where
\be
w = {P^{(CM)}_N(\lambda )\over P^{(CM)}_N(\lambda - m)}
\ \ \ \ \ \ \ \ \
dS = \lambda{dw\over w}.
\label{triCa}
\ee
Five dimensional gauge theories whose $D=4$ reductions have adjoint matter
correspond to the trigonometric Ruijsenaars-Schneider system, where now
\be
w=\frac{P^{(RS)}_N(\lambda)}
{P^{(RS)}_N(\lambda e^{-2i\epsilon})}
\ \ \ \ \ \ \ \ \
dS = \log\lambda {dw\over w}.
\label{triRu}
\ee
In each of these cases $P^{(CM)}_N$ and $P^{(RS)}_N$ are appropriate polynomials
that we shall further describe in due course. It is easy to see that
(the perturbative) spectra are given by the general formula \cite{BMMM2}
\be
M = a_{ij} \oplus {\pi n\over R} \oplus {\epsilon +\pi n\over R},
\qquad
n\in {\bf Z}.
\label{spectrum}
\ee
In addition to the Higgs part $a_{ij}$ this contains the Kaluza-Klein (KK)
modes $\pi n\over R$ and the KK modes for fields with ``$\epsilon$-shifted"
boundary conditions. The effective couplings are defined by
almost the same formula as (\ref{effcharge})
\be
T_{ij}^{\rm pert} = {1\over 2\pi i}\sum_{(\rm masses\ M)_{ij}}
\log {M^2\over\Lambda ^2}
\ee
i.e. by the sum of all the logarithms of spectrum (\ref{spectrum}) giving
contribution for a particular 1-loop diagram for $T_{ij}$.

{\em A priori} nothing so simple can be said about the spectrum and structure
of the theory at strong couplings.
However the structure of the ``strongly-coupled" phase is, as we will see
below using the correspondence with integrable systems,
similar in many respects to the weak-coupling regime, and may
be described by explicitly computable expressions.

\subsection{Generalities: finite-gap or Hitchin systems}

Let us now consider the general setting.
Our analysis so far has been built upon a (polynomial or rational)
relation $R(\lambda, w)=0$,
a differential $dS$ and BPS data given by integrating $dS$ over various
contours. We will now consider $R(\lambda, w)=0$ as the equation describing
the spectral curve of an integrable system. The differential, contours
and symplectic form may be canonically described in terms of this integrable
system. At heart will be the choice of integrable system. For the SW theories
under discussion these are given by finite gap integrable systems of
particles of a particular kind, the so-called
Hitchin systems \cite{Hi,GoNe}. We shall now describe
some of these general features and give explicit descriptions of the Hitchin
systems of relevance to us. The remarkable feature of the SW integrable
systems correspondence is that non-perturbative aspects of the field theory
are incorporated into this construction \cite{don, SWCal}.

Replacing the relation $R(\lambda, w)=0$ of (\ref{polyn}), (\ref{triCa}) and
(\ref{triRu}) one now has the Lax equation of a spectral curve,
\be
\label{laxcu}
\det (\lambda -  {\cal L}(z)) = 0.
\ee
Here the Lax operator ${\cal L}(z)$ is a matrix, defined on some {\em base}
curve $\Sigma_0$ on which the spectral parameter $z$ lies. For us this base
curve is usually a torus (for elliptic models) or a sphere with
punctures (for rational or trigonometric
\footnote{In perturbative examples it is parameterized by $w=e^{z}$.
} models). The two further ingredients were the generating differential $dS$
and the contours $C_{ij}$. The generating differential can now be defined by
\be
\label{dS}
dS = \lambda dz
\\
\delta_{\rm moduli} dS = {\rm holomorphic\  differential}.
\ee
The {\em action} variables are given by the Seiberg-Witten contour
integrals over half of the independent contours
\be
{\bf a} = \oint_{\bf A}dS,
\ee
or
\be
{\bf a}^D = \oint_{\bf B}dS.
\ee
where ${\bf A}=\{ A_1,\dots,A_g\}$ and ${\bf B}=\{ B_1,\dots,B_g\}$ is
a standard homology basis of $H^1(\Sigma)$.

The holomorphic variation in (\ref{dS}) is crucial to the definition of
the symplectic form
\be
\label{omega}
\Omega = \left.\delta dS\right|_{\bgamma} = d{\bf a}\wedge d{\bf z}(\bgamma)
= d{\bf p}\wedge d{\bf q}.
\ee
The variation here is to be computed at the divisor
${\bgamma} = \{\gamma_1,\dots,\gamma_g\}$ of the poles of the Baker-Akhiezer
function (which is determined by ${\cal L}(z)$)
which are co-ordinates on $\Sigma^{g}$.
Let $\{d\omega_i\} $ be the the set of {\em canonically} normalised holomorphic
differentials $\oint_{A_i}d\omega_j=\delta_{ij}$.
The variation of $dS$ is understood as a total external derivative on
${\rm Moduli}_g \ltimes \Sigma\sp g$,
\be
\label{dds}
\delta dS = \delta (\lambda dz) = \delta_{\rm moduli}\lambda\wedge dz +
d\lambda\wedge dz.
\ee
For this one must choose a {\em connection} on the bundle over moduli space
\cite{KriPho}
such that $\delta_{\rm moduli} z = 0$, i.e. the parameter on the base curve
$z$ is covariantly constant. In practice this means that in the equation on two
variables $\lambda$ and $z$ which defines the spectral curve $\Sigma $, we
consider $z$ as an independent variable, with the variable $\lambda$
depending on moduli through eq.~(\ref{laxcu}). Since on a one-dimensional
curve $\Sigma $ for any differentials one has
$d\lambda\wedge dz=0$, from (\ref{dds}) we finally get
\be
\delta dS = \delta_{\rm moduli}\lambda\wedge dz =
\delta a_i \wedge {\d\lambda\over\d a_i}dz =
\delta a_i \wedge {\d dS\over\d a_i} =
\delta a_i \wedge d\omega_i.
\ee
Finally we introduce co-ordinates ${\bf z}$ on the Jacobian by
\be
\label{angles}
z_i = \int_{\gamma_0}^{\bgamma}d\omega_i + z_i^{(0)} \equiv
\sum_{j=1}^g \int_{\gamma_0}^{\gamma_j}d\omega_i + z_i^{(0)},
\ee
with $z_i^{(0)}= z_i^{(0)}(\gamma_0)$.
Then
\be
\Omega = \sum_{k=1}^g\delta dS(\gamma_k) =
\delta a_i \wedge \sum_{k=1}^gd\omega_i(\gamma_k)
\stackrel{(\ref{angles})}{=}
\delta a_i\wedge\delta z_i.
\ee

Our discussion thus far is general, predicated only upon a Lax operator and its
attendant Baker-Akhiezer function. We must now describe the Lax operators
for the class of models under consideration. These will be particular
examples of Hitchin systems.  For these the Lax operator in (\ref{laxcu})
${\cal L} (z)\equiv \bPhi $ is considered as a meromorphic matrix-valued
function (or better, a 1-differential) on the base curve
$\Sigma_0$ \cite{DKN,Hi,GoNe} satisfying
\be
{\bar\partial}\Phi + [{\bar A},\Phi ]=\sum_{\alpha}J^{(\alpha )}
\delta^{(2)}(P-P_{\alpha}).
\label{gauss}
\ee
The right hand side serve as sources, and $J^{(\alpha )}$ are matrices whose
structure (given shortly for a single puncture) is particularly simple for
the $su(n)$ theories.
Thus $\bPhi $ is {\em holomorphic} in the complex structure determined by
$\bar{A}$ on the punctured curve $\Sigma_0 /\{P_{\alpha}\}$.
The invariants of $\bar A$ can be thought of as co-ordinates
(one commuting set of variables) while the invariants of $\Phi$ as
hamiltonians (another commuting set of variables) of an integrable system.
The most general features of these ``holomorphic" finite-gap \cite{DKN}
or Hitchin \cite{Hi,GoNe} systems are:

\begin{itemize}
\item The spectral curve $\Sigma_g$ covers some base curve $\Sigma_0$ (typically
of genus $g_0=0,1$) and the $g$ moduli of the cover are viewed as
distinguished with the moduli of the base curve viewed as ``fixed" or
``external" parameters". This set of distinguished moduli are to be the
constants of motion of our integrable system.

\item In the general context of finite-gap integrable systems one may view this
set-up in a different way. A generic complex curve depends on $3g-3$ parameters
describing the complex structure.
The integrable system is described \cite{DKN,KriPho} by two meromorphic
differentials $d\lambda $ and $dz$ on the curve.
By adding holomorphic differentials these may be taken as having vanishing
$\bf A$ periods. Fixing the $\bf B$ periods specifies the integrable system.
Loosely, $d\lambda $ and $dz$ are only defined up to multiples, and fixing
the $\bf B$ periods of $d\lambda $ gives $g-1$ constraints (taking account of
the scaling freedom), while the $\bf B$ periods of $dz$ give a further
$g-2$ constraints (now allowing $dz\rightarrow \alpha dz +\beta d\lambda $).
Altogether we come to a system with $(3g-3)-(g-1)-(g-2)=g$ parameters, the
genus of $\Sigma$.

\item This construction (when $g_0=0,1$) implies certain linear
relations on the period matrix $T_{ij}$:
\be
\label{Tconstr}
\sum_j T_{ij} = \tau = {\rm fixed}.
\ee
This comes from the possibility to choose the homology basis of $\Sigma $ such
that the ${\bf A}$ and ${\bf B}$ cycles ``project" to the $A$ and $B$
cycles on the base torus $\Sigma_0$. (A rational base curve can be thought of
as degenerated torus.) Then
\be
\label{suma}
\sum_i a_i = \sum_i \oint_{A_i}\lambda dz = \oint_A
\left(\sum\lambda_i\right)dz = \sum {\cal L}_{ii}\oint_A dz = h_1\oint_A dz
= h_1,
\ee
and
\be
\label{sumb}
\sum_i a^D_i = \sum_i \oint_{B_i}\lambda dz = \oint_B
\left(\sum\lambda_i\right)dz = \sum {\cal L}_{ii}\oint_B dz = h_1\oint_B dz
= h_1\tau.
\ee
Thus, for any $j$,
\be
\sum_i T_{ij} = \sum_i {\d a^D_i\over\d a_j} = {\d \over\d a_j}\sum_i a^D_i\
\stackrel{(\ref{sumb})}{=}{\d h_1\over\d a_j}\ \tau\
\stackrel{(\ref{suma})}{=}\ \tau
\ee
giving (\ref{Tconstr}).

\end{itemize}
These properties we will be crucial in our discussion of the {\em
tau-functions} (basically the Riemann theta-functions on the
corresponding spectral curves $\Sigma$) for these finite-gap integrable
systems.

Let us now give three particular examples of the Hitchin integrable models
arising in the context of SW theory: the Toda chain, the elliptic
Calogero-Moser model and the Ruijsennars-Schneider model. These are
integrable systems we encountered in the perturbative discussion earlier.

{\bf Toda chain}. The Toda chain system \cite{Toda} is a system of
$N$ particles with nearest neighbour exponential interactions:
\be\label{Todaeq}
\frac{\partial  q_i}{\partial t} = p_i \ \ \ \ \
\frac{\partial p_i}{\partial t} = e^{ q_{i+1} - q_i}-
e^{ q_i- q_{i-1}}.
\ee
It is an integrable system, with $N$
Poisson-commuting Hamiltonians, $h_1 = \sum p_i$ = P, $h_2 =
\sum\left(\frac{1}{2}p_i^2 + e^{ q_i- q_{i-1}}\right) = E$, etc.
We may have either an open chain ($q_0=-\infty$, $q_{N+1}=\infty$)
or a periodic system (of ``period" $N$: $ q_{i+N} =  q_i$ and $p_{i+N} = p_i$).
The periodic problem can be derived by reduction of the infinite-dimensional
system of particles on the line by aid of two commuting
operators: the Lax operator ${\cal L}$ (for the auxiliary linear problem for
(\ref{Todaeq}))
\be\label{laxtoda}
\lambda\Psi _n =
\sum _k {\cal L}_{nk}\Psi _k =
e^{{1\over 2}( q_{n+1}- q_n)}\Psi _{n+1} + p_n\Psi _n +
e^{{1\over 2}( q_n- q_{n-1})}\Psi _{n-1}
\ee
and a second operator, the {\it monodromy} or shift operator in the
discrete variable (the particle number)
\be\label{T-op}
T q_n =  q_{n+N}\ \ \ \ \ \ Tp_n = p_{n+N}\ \ \ \ \ \ \ \
T\Psi_n = \Psi_{n+N}.
\ee
The existence of a common spectrum for these two operators
\be\label{spec}
{\cal L}\Psi = \lambda\Psi \ \ \ \ \ T\Psi = w\Psi\ \ \ \ \ \ [{\cal L},T]=0
\ee
means that there is a relation between them ${\cal P}({\cal L},T) = 0$.
This in fact is the equation of the spectral curve $\Sigma$:
${\cal P}(\lambda,w) = 0$.

One way to get explicit form of the spectral curve equation is to rewrite the
Lax operator (\ref{laxtoda}) in the basis of the $T$-operator
eigenfunctions.\footnote{
If we had chosen to work instead with ${\cal L}$, which is a
{\em second}-order difference operator, we would come to the
Faddeev-Takhtajan $2\times 2$ formalism of Toda chains.  }
This then becomes a $N\times N$ matrix,
\be\label{LaxTC}
{\cal L} = {\cal L}^{TC}(w) = {\bf p\cdot H} +
\sum _{{\rm simple}\ \balpha}e^{\balpha\cdot \Bf q}\left( E_{\alpha}
+ E_{-\alpha}\right) + w^{-1}e^{-\balpha _0\cdot \Bf q}E_{-\alpha _0} +
w e^{-\balpha _0\cdot \Bf q}E_{\alpha _0} \\
\\ =
\left(\begin{array}{ccccc}
 p_1 & e^{{1\over 2}( q_2- q_1)} & 0 &\ldots & we^{{1\over 2}
( q_1- q_{N})}
\\
e^{{1\over 2}( q_2- q_1)} & p_2 & e^{{1\over 2}( q_3 -  q_2)} &
\ldots & 0\\
0 & e^{{1\over 2}( q_3- q_2)} & p_3 &\ldots & 0 \\
\vdots & &  &\ddots & \vdots\\
\frac{1}{w}e^{{1\over 2}( q_1- q_{N})} & 0 & 0 &\ldots & p_{N}
\end{array} \right).
\ee
Here the sum is over the simple roots, and
$\balpha_0 = - \sum _{{\rm simple}\ \balpha}\balpha$ is minus the highest root.
The construction depends explicitly on
the eigenvalue $w$ of the shift operator (\ref{T-op}). This is the
spectral parameter which is defined on the cylinder
$\Sigma_0={\bf CP}\sp1\setminus\{0,\infty \}$.
The eigenvalues of the Lax operator (\ref{LaxTC}) are defined from the spectral
equation
\be\label{SpeC}
{\cal P}(\lambda,w) = \det_{N\times N}\left(\lambda - {\cal L}^{TC}(w)
\right) = 0.
\ee
Substituting the explicit expression (\ref{LaxTC}) into (\ref{SpeC}),
one obtains
\be\label{fsc-Toda}
{\cal P}(\lambda,w) =  P_{N}(\lambda ) - w - \frac{1}{w} = 0,
\qquad
P_N(\lambda) = \lambda^N + \sum_{k=1}^{N}
(-1)\sp{k}\,h_k\lambda^{N-k},
\ee
which for $N=2$ is
\be\label{Toda2}
w+\frac{1}{w}=\lambda\sp2-(p_1+p_2)\lambda+
p_1 p_2 -(e\sp{q_2-q_1}+e\sp{q_1-q_2}).
\ee
The generating 1-form $dS = \lambda{dw\over w}$ indeed satisfies (\ref{dS})
\be
\delta_{\rm moduli}dS \equiv \left.\delta_{\rm moduli}dS \right|_{w=const}
= \left(\delta_{\rm moduli}\lambda\right){dw\over w} =
{\sum\lambda^k \delta h_k \over P_N'(\lambda)}{dw\over w} =
\sum{\lambda^kd\lambda\over y}\delta h_k = {\rm holomorphic}.
\label{hol}
\ee
where
\be
\label{defytoda}
y^2 = \left(w - {1\over w}\right)^2 = P_N^2(\lambda ) - 4,
\qquad
P_N'(\lambda ) \equiv\left. {\d P_N\over\d\lambda }\right|_{h_k =const}
\ee
By the gauge transformation $U_{ij} = v^i\delta _{ij}$ with $ w\equiv v^{N}$
the Lax operator (\ref{LaxTC}) can be brought to another familiar form
\begin{equation}
\label{LaxTCHi}
\begin{array}{rl}
{\cal L}^{TC}(w) &\rightarrow \tilde{\cal L}^{TC}(v) = U^{-1}{\cal L}^{TC}(w)U
 \\
&= {\bf p\cdot H} + v^{-1}\left( e^{-\balpha _0\cdot \Bf q}E_{-\alpha _0} +
\sum _{{\rm simple}\ \balpha}e^{\balpha\cdot \Bf q}E_{\alpha}\right)+
 v\left( e^{-\balpha _0\cdot \Bf q}E_{\alpha _0} +
\sum _{{\rm simple}\ \balpha}e^{\balpha\cdot \Bf q}E_{-\alpha}\right)
\\
\\
&= \left(\begin{array}{ccccc}
 p_1 & \frac{1}{v}e^{{1\over 2}( q_2- q_1)} & 0 & \ldots
& ve^{{1\over 2}( q_1- q_{N})}\\
ve^{{1\over 2}( q_2- q_1)} & p_2
& \frac{1}{v}e^{{1\over 2}( q_3 -  q_2)} & \ldots & 0\\
0 & ve^{{1\over 2}( q_3- q_2)} & p_3 & \ldots & 0 \\
\vdots & &  & \ddots & \vdots \\
\frac{1}{v}e^{{1\over 2}( q_1- q_{N})} & 0 & 0 & \ldots & p_{N}
\end{array} \right) .
\end{array}
% \\ U_{ij} = v^i\delta _{ij} \ \ \ \ \ w\equiv v^{N}
\end{equation}
In this form \cite{M97} it clearly satisfies the $\bar{\partial}$-equation
(\ref{gauss}) on a cylinder with {\em trivial} gauge connection
\be\label{hito}
\bar\partial _v \tilde{\cal L}^{TC}(v) = {\partial\over\d\bar v}
\tilde{\cal L}^{TC}(v) =
\left( e^{-\balpha _0\cdot \Bf q}E_{-\alpha _0} +
\sum _{{\rm simple}\ \balpha}e^{\balpha\cdot \Bf q}E_{\alpha}\right)
\delta (P_0) \\ -
\left( e^{-\balpha _0\cdot \Bf q}E_{\alpha _0} +
\sum _{{\rm simple}\ \balpha}e^{\balpha\cdot \Bf q}E_{-\alpha}\right)
\delta (P_{\infty}).
\ee
The Toda chain coupling $\Lambda $ (equal to the mass scale $\Lambda_{QCD}$
in the SW theory) can be restored by rescaling $\lambda\to\lambda/\Lambda$,
$w\to w/\Lambda^N$ and the hamiltonians $h_k\to{h_k\over\Lambda^k}$. Then
equation (\ref{fsc-Toda}) acquires the form of
\be
\label{todacu}
w + {\Lambda^{2N}\over w} = P_N(\lambda) = \lambda^N + \sum_{k=1}^{N}
(-1)\sp{k}\,h_k\lambda\sp{N-k}
\ee
proposed in the context of SW theory \cite{SW+}.

{\bf Elliptic Calogero-Moser model}. The $N\times N$ matrix Lax operator
(\ref{LaxTCHi}) (and, therefore, (\ref{LaxTC})) can
be thought of as a ``degenerate" case of the Lax operator for the $N$-particle
Calogero-Moser system \cite{KriCal}
\begin{equation}
\label{LaxCal}
\begin{array}{rl}
{\cal L}^{CM}(z) &=
{\bf p\cdot H} + \sum_{\balpha}F({\bf q\cdot\balpha}|z) E_{\balpha}  \\
\\
&= \left(\begin{array}{cccc}
 p_1 & F(q_1-q_2|z) & \ldots & F(q_1 - q_{N}|z)\\
F(q_2-q_1|z) & p_2 & \ldots & F(q_2-q_{N}|z)\\
\vdots & & \ddots  & \vdots \\
F(q_{N}-q_1|z) &
F(q_{N}-q_2|z)& \ldots &p_{N}
\end{array} \right).
\end{array}
\end{equation}
The sum here is now over all of the roots and the
base curve $\Sigma_0$ is a torus instead of a cylinder.
The matrix elements
\be
\label{LaxCalma}
F(q|z) = i m\frac{\sigma(q+z)}{\sigma(q)\sigma(z)}
\ee
are defined in terms of the Weierstrass sigma-functions $\sigma(z)$.
Equivalently $\sigma (z) =
2\omega e^{\eta z^2\over 2\omega}{\theta_\ast (z)\over\theta'_\ast(0)}$ where
$\theta_{\ast}(z)\equiv\theta[{1\atop1}](z)\equiv\theta_1(z)$ is the (only)
odd Jacobi theta-function. The modulus of the elliptic curve $\Sigma_0$ is
$\tau $ and we have the marked point
$z=0$ at which $x=\wp(z)=\infty$, $y=\2\wp'(z)=\infty$.
The canonical holomorphic 1-differential on $\Sigma_0$ is $dz = \frac{dx}{2y}$.
The Lax operator (\ref{LaxCal}) corresponds to a completely
integrable system with Hamiltonians $h_1 = \sum_i p_i$, $h_2 =
\sum_{i<j}(p_ip_j - m^2\wp(q_i-q_j))$, $h_3 = \sum_{i<j<k}(p_ip_jp_k+\dots)$,
etc.

From (\ref{laxcu}), (\ref{LaxCal}) it follows that the spectral curve
$\Sigma^{CM}$ for the $N$-particle Calogero-Moser system
\be\label{fscCal}
\det_{N\times N}\left(\lambda - {\cal L}^{CM}(z)\right) = 0
\ee
covers $N$ times the base elliptic curve. These spectral curves are very
special: in general the genus
of a curve defined by an $N\times N$ matrix grows as $N^2$ while the
curve (\ref{fscCal}) has genus $g = N$. For $N=2$ we have
\be
\label{CM2}
0=\lambda\sp2-(p_1+p_2)\lambda+ p_1 p_2 -
m\sp2 \left(\wp(q_2-q_1)-\wp(z)\right).
\ee
The BPS masses ${\bf a}$ and ${\bf a}\sp{D}$ are now related to the periods
of generating 1-differential
\be\label{dSCal}
dS^{CM} = 2\lambda dz = \lambda{dx\over y}
\ee
along the non-contractable contours on $\Sigma^{CM}$.

In order to recover the Toda-chain system, one has to take the double-scaling
limit \cite{Ino}, when $m$ and $-i\tau$ both go
to infinity and
\be
\label{dscaling}
q_i-q_j\rightarrow {\2}(i-j)\log m +(q_i-q_j)
\ee
so that the dimensionless coupling $\tau$ gets
substituted by a dimensionful parameter $\Lambda^{N} \sim
m^{N}e^{i\pi\tau}$. The idea is to separate the pairwise interacting particles
far away from each other and to adjust the coupling constant simultaneously
in such a way, that only the interaction of nearest particles survives (and
turns in an exponential). This limit is described in more detail in Appendix A.
In this limit, the elliptic curve degenerates into
a cylinder with coordinate $w = e^{z}e^{i\pi\tau}$ so that $dz\to{dw\over
w}$ and
\be
dS^{CM} \rightarrow dS^{TC} = \lambda\frac{dw}{w}.
\ee
The Lax operator of the Calogero system turns into that of the
$N$-periodic Toda chain (\ref{LaxTC}):
\be
{\cal L}^{CM}(z )dz \rightarrow {\cal L}^{TC}(w)\frac{dw}{w}
\ee
and the spectral curve acquires the form (\ref{SpeC}).

One further remark is in order concerning the dependence of the elliptic
Calogero-Moser model on the coupling constant, or equivalently
the mass of the adjoint hypermultiplet $m$. The exact
equivalence between the Calogero-Moser and KP theories is usually considered
only when $m^2 = 2$ (or at least $m^2 = n(n+1)$ for integer $n$). This
restriction is however only essential when we consider the ``Lax" equation
for the ``first" KP time, the $x$-variable,
\be
\left(- {\d^2\over\d x^2} + m^2\sum \wp (x-q_i)\right)\Psi = E\Psi.
\ee
This
has solution with simple poles $\Psi\sim\sum {c_i\over x-q_i} + \dots$ only
when $m^2 = 2$ (the condition of cancellation of the highest pole). However,
for the Calogero-Moser equations themselves (${d^2q_i\over dt^2} =
m^2\sum_{i<j}\wp'(q_{ij})$ and the ``higher" hamiltonian equations)
there is no such restriction on the coupling constant. One can
always set $m^2=2$ by a simple rescaling of the time-variables $t\to mt$.
The point is that in achieving the canonical forms of the KP and KdV
equations all of the possible scalings have already been employed: consequently
the coupling constant is restricted.

{\bf Calogero-Moser spectral curve from the elliptic Ruijsenaars}.
The easiest and most general way to look at Seiberg-Witten theories with
adjoint matter is to consider the elliptic Ruijsenaars-Schneider model
\cite{BMMM2}.
The Lax operator of the elliptic Ruijsenaars model has the form
\cite{Rui}
\be
{\cal L}^R_{ij} =
e^{P_i} \frac{\sigma (q_{ij}+z+\epsilon)\sigma (\epsilon)}
{\sigma(q_{ij}+\epsilon)\sigma(z+\epsilon)},
\qquad
e^{P_i} = e^{p_i}
\prod_{k\neq i}\sigma(\epsilon) \sqrt{\wp(\epsilon) - \wp (q_{ik})}.
\label{RL}
\ee
In the trigonometric limit this turns into
\be
{\cal L}^{TR}_{ij} =
e^{P_i} \frac{\sinh (q_{ij}+z+\epsilon)\sinh (\epsilon)}
{\sinh(q_{ij}+\epsilon)\sinh(z+\epsilon)},\qquad
e^{P_i} = e^{p_i}
\prod_{k\neq i} \sqrt{1 - {\sinh^2\epsilon\over\sinh^2(q_{ik})}}.
\label{TRL}
\ee
Introducing $\nu_i = e^{2q_i}$, $\zeta = e^{2z}$ and $q = e^{2\epsilon}$ one
finds that
\be
{\cal L}^{TR}_{ij} =
e^{P_i} \frac{q\zeta\nu_i - \nu_j}
{q\nu_i - \nu_j}{q-1\over q\zeta-1} \stackreb{\zeta\to\infty}{=}
(q-1){e^{P_i}\nu_i\over q\nu_i - \nu_j} + {\cal O}\left({1\over\zeta}\right),
\quad\quad
e^{P_i} = e^{p_i}
\prod_{k\neq i} {\sqrt{q\nu_i-\nu_j}\sqrt{\nu_i-q\nu_j}
\over \nu_i - \nu_j}.
\label{TRL1}
\ee
Often only the leading term in (\ref{TRL1}) is taken as an expression for
the Lax operator of the trigonometric Ruijsenaars system.

Using (\ref{RL}) the spectral curve equation for the elliptic
Ruijsenaars can be written as
\footnote{This is a particular case of generic determinant formulas
(\ref{fr11}), considered in more detail below.}
\begin{equation}
\det_{N\times N}\left(\lambda - {\cal L}^R(z)\right) =
\sum_{k=0}^N\lambda^{N-k}(-)^k\sum_{|I|=k}\prod_{i\in I}e^{P_i}\left.
\det_{k\times k}\frac{\sigma (q_{ij}+z+\epsilon)\sigma (\epsilon)}
{\sigma(q_{ij}+\epsilon)\sigma(z+\epsilon)}\right|_{i,j\in I}
\label{Rsc}
\end{equation}
where, using Wick's theorem for the correlation functions of free
fermions \cite{Wick,Rui,BMMM2}, for the determinants one
gets
\begin{equation}
\label{dets1}
\begin{array}{rl}
\det_{k\times k}&\frac{\sigma (q_{ij}+z+\epsilon)\sigma (\epsilon)}
{\sigma(q_{ij}+\epsilon)\sigma(z+\epsilon)} =
\det_{k\times k}\frac{\theta_\ast (q_{ij}+z+\epsilon)\theta_\ast (\epsilon)}
{\theta_\ast(q_{ij}+\epsilon)\theta_\ast(z+\epsilon)}\ \stackreb{{\tilde
q}_j=q_j-\epsilon}{=}\
\det_{k\times k}\frac{\theta_\ast (q_i-{\tilde q}_j+z)}
{\theta_\ast(q_i - {\tilde q}_j)\theta_\ast(z)}\ {\theta_\ast(z)\theta_\ast (\epsilon)\over
\theta_\ast(z+\epsilon)}
\\
&= \left({\theta_\ast(z)\theta_\ast (\epsilon)\over\theta_\ast(z+\epsilon)}\right)^k
\det_{k\times k}\langle\psi(q_i){\tilde\psi}(q_j)\rangle_z =
\left({\theta_\ast(z)\theta_\ast (\epsilon)\over\theta_\ast(z+\epsilon)}\right)^k
\langle\prod_{i\in I}\psi(q_i)\prod_{j\in I}{\tilde\psi}(q_j)\rangle_z
\\
&= \left({\theta_\ast(z)\theta_\ast (\epsilon)\over\theta_\ast(z+\epsilon)}\right)^k
{\prod_{i<j}\theta_\ast(q_i-q_j)\prod_{i<j}\theta_\ast({\tilde q}_i-{\tilde q}_j)
\over\prod_{i,j}\theta_\ast(q_i-{\tilde q}_j)}\ {\theta_\ast(\sum q_i - \sum {\tilde
q}_i+z)\over\theta_\ast(z)}
\\
&= \left({\theta_\ast(z)\over\theta_\ast(z+\epsilon)}\right)^k
{\prod_{i<j}\theta_\ast(q_{ij})^2\over\prod_{i\neq j}\theta_\ast(q_{ij}+\epsilon)}
{\theta_\ast(z+k\epsilon)\over\theta_\ast(z)}.
\end{array}
\end{equation}
Using (\ref{dets1}) and introducing
$\tilde\lambda\equiv\lambda{\theta_\ast(z+\epsilon)\over\theta_\ast(z)}$, one finally
gets for (\ref{Rsc})
\be
\label{rucu}
\det_{N\times N}\left(\lambda - {\cal L}^R(z)\right) =
\left({\theta_\ast(z)\over\theta_\ast(z+\epsilon)}\right)^N
\left[\sum_{k=0}^N{\tilde\lambda}^{N-k}h_k
{\theta_\ast(z+k\epsilon)\over\theta_\ast(z)}\right]=0
\ee
where the hamiltonians $h_k$ are
\begin{equation}
\label{ruham}
\begin{array}{rl}
(-)^kh_k& \equiv \sum_{|I|=k}\prod_{i\in I}e^{P_i}
{\prod_{i<j\in I}\theta_\ast(q_{ij})^2\over\prod_{i\neq j\in I}
\theta_\ast(q_{ij}+\epsilon)} =
\sum_{|I|=k}\prod_{i\in I}e^{P_i}
\prod_{i<j\in I}{\theta_\ast(q_{ij})^2\over
\theta_\ast(q_{ij}+\epsilon)\theta_\ast(-q_{ij}+\epsilon)}
\\
&= \sum_{|I|=k}\prod_{i\in I}e^{P_i}
\prod_{i<j\in I}{1\over\theta_\ast(\epsilon)^2}\ {1\over\wp(\epsilon)-\wp(q_{ij})}
\end{array}
\end{equation}
where we used $${\theta_\ast(q+\epsilon)\theta_\ast(q-\epsilon)\over
\theta_\ast(q)^2\theta_\ast(\epsilon)^2} = {\sigma(q+\epsilon)\sigma(q-\epsilon)\over
\sigma(q)^2\sigma(\epsilon)^2} = \wp(\epsilon)-\wp(q).$$

For $N=2$ we obtain
\begin{equation}
\label{RS2}
\begin{array}{rl}
\det_{2\times 2}\left(\lambda - {\cal L}^R(z)\right) &=
\lambda^2 -\lambda\left(e\sp{P_1}+e\sp{P_2}\right)+e\sp{P_1+P_2}
\frac{\wp(\epsilon)-\wp(z+\epsilon)}{\wp(\epsilon)-\wp(q_{12})}\\
&= \lambda^2 - \lambda\sigma(\epsilon)\left(e^{p_1}+
e^{p_2}\right)\sqrt{\wp(\epsilon)-\wp(q_{12})} +
e^{p_1+p_2}\sigma^2(\epsilon)\left(\wp(\epsilon)-\wp(z+\epsilon)\right) .
\end{array}
\end{equation}

The spectral curve of the elliptic Calogero-Moser model
(\ref{LaxCal}) arises in
the $\epsilon\to 0$ limit of (\ref{rucu}). Indeed, in the limit
$\epsilon=mR$, $p_i\to Rp_i$ and $R\to 0$, one recovers from (\ref{RL})
\be
{\cal L}_{ii}^R = e^{P_i} = e^{Rp_i}
\prod_{k\neq i}\sigma(mR) \sqrt{\wp(mR)-\wp (q_{ik})} =1+Rp_i +{\cal O}(R^2),
\\
\\
{\cal L}_{ij}^R = e^{Rp_i}
\prod_{k\neq i}\sigma(mR) \sqrt{\wp(mR)-\wp (q_{ik})}\
 \frac{\sigma (q_{ij}+z+mR)\sigma (mR)}
{\sigma(q_{ij}+mR)\sigma(z+mR)}
= R\cdot m\frac{\sigma (q_{ij}+z)}{\sigma(q_{ij})\sigma(z)}+{\cal O}(R^2)
\quad i\neq j .
\ee
Thus
\begin{equation}
{\cal L}_{ij}^R = \delta_{ij} + R\left(p_i\delta_{ij} +
m(1-\delta_{ij})\frac{\sigma (q_{ij}+z)}{\sigma(q_{ij})\sigma(z)}\right)
+{\cal O}(R^2) =
\delta_{ij} + R{\cal L}_{ij}^{CM} +{\cal O}(R^2)
\end{equation}
and one obtains
\begin{equation}
\begin{array}{rl}
\det_{N\times N}\left(\lambda - {\cal L}^R(z)\right) &=
\det_{N\times N}\left(\lambda - 1 - R{\cal L}^{CM}(z) + {\cal O}(R^2)\right) \\
&= R^N\det_{N\times N}\left({\lambda-1\over R} - {\cal L}^{CM}(z)\right) +
{\cal O}(R^{N+1}).
\end{array}
\end{equation}
Provided the Ruijsenaars parameter scales as
${\lambda-1\over R}\to\lambda\sp{CM}$
then in the limit $R\to 0$ we obtain exactly the spectral equation
(\ref{fscCal}).

Lets consider these for the $N=2$ example. In this case one has from (\ref{RS2})
\begin{equation}
\label{run2ca}
\begin{array}{rl}
\det_{2\times 2}\left(\lambda - {\cal L}^R(z)\right)&
= (\lambda-1)^2-m R(\lambda-1)(p_1+p_2) \\ &+(m R)^2
\left(\lambda\wp(q_{12}) - \2\lambda(p_1^2+p_2^2) + \2(p_1+p_2)^2 - \wp(z)
\right) + {\cal O}(R^3)
\end{array}
\end{equation}
After the substitutions $\lambda-1\to R\lambda'$,
one finally gets from (\ref{run2ca})
\be
\det_{2\times 2}\left(R\lambda' +1 - {\cal L}^R(z)\right)\stackreb{R\to 0}{=}
R^2\left(\lambda'^2 - \lambda' (p_1+p_2) + p_1p_2 + m^2\wp(q_{12}) -
m^2\wp (z)\right) + {\cal O}(R^3)
\ee
which is the $N=2$ elliptic Calogero-Moser curve up to $R^{N+1}=R^3$ terms. On
the other hand the $N=2$ spectral curve (\ref{run2ca}) can be rewritten in
the form (\ref{rucu})
\be
\label{run2dph}
\det_{2\times 2}\left(\lambda - {\cal L}^R(z)\right) =
\left({\sigma(z)\over\sigma(z+\epsilon)}\right)^2\left(
{\tilde\lambda}^2 - \tilde\lambda{\sigma(z+\epsilon)\over\sigma(z)}\,
h_1\sp{R} + {\sigma(z+2\epsilon)\over\sigma(z)}e\sp{p_1+p_2}\right)
\ee
where
\be
\tilde\lambda = \lambda{\sigma(z+\epsilon)\over\sigma(z)},\qquad
h_1\sp{R} = \left(e^{p_1}+e^{p_2}\right)\sigma(\epsilon)
\sqrt{\wp(\epsilon)-\wp(q_{12})}.
\ee
With $p_i\to \epsilon p_i$ then in the limit $\epsilon=mR\to 0$
this may be expressed as
\begin{equation}
\begin{array}{rl}
{\tilde\lambda}^2 - \tilde\lambda{\sigma(z+\epsilon)\over\sigma(z)}\,
h_1\sp{R} + &{\sigma(z+2\epsilon)\over\sigma(z)}e\sp{p_1+p_2}=
(\tilde\lambda-1)^2 - \epsilon(\tilde\lambda-1)(h_1\sp{CM}+2
{\sigma'(z)\over\sigma(z)})\\& + \epsilon^2\left(\tilde\lambda h_2^{CM} +
(2-\tilde\lambda)({\sigma''(z)\over\sigma(z)}-{\sigma'(z)\over\sigma(z)}
h_1\sp{CM})
+\2(\tilde\lambda-1)(h_1\sp{CM})\sp{2}\right) + {\cal O}(\epsilon^3).
\end{array}
\end{equation}
Here we have used that
$h_1^R = 2 +\epsilon h_1\sp{CM}+\epsilon^2\left(
\2(h_1\sp{CM})\sp{2}-h_2^{CM}\right).$
Finally after the substitution $\tilde\lambda-1\to\epsilon\lambda'$, one
arrives at
\begin{equation}
\label{dhph2}
\begin{array}{rl}
{\tilde\lambda}^2 - \tilde\lambda{\sigma(z+\epsilon)\over\sigma(z)}\,
h_1\sp{R} + {\sigma(z+2\epsilon)\over\sigma(z)}e\sp{p_1+p_2}
&=
\epsilon^2\left(\lambda'^2 -\lambda'h_1\sp{CM} +h_2^{CM}
+(h_1\sp{CM}- 2\lambda')
{\sigma'(z)\over\sigma(z)} + {\sigma''(z)\over\sigma(z)}\right)
+ {\cal O}(\epsilon^3)\\
&=
\epsilon^2\ {\sigma\left(z-{\d\over\d\lambda'}\right)\over\sigma(z)}
\left(\lambda'^2 -\lambda'h_1\sp{CM} +h_2^{CM}\right) + {\cal O}(\epsilon^3).
\end{array}
\end{equation}
One can also rewrite (\ref{dhph2}) as
\begin{equation}
\label{cdhph2}
\begin{array}{rl}
\lambda'^2 &-\lambda'h_1\sp{CM} +h_2^{CM}
+(h_1\sp{CM}- 2\lambda')
{\sigma'(z)\over\sigma(z)} + {\sigma''(z)\over\sigma(z)}\\
&=\left(\lambda' - {\sigma'(z)\over\sigma(z)}\right)^2 -
\left(\lambda' - {\sigma'(z)\over\sigma(z)}\right)h_1\sp{CM}
-{\sigma'(z)^2\over\sigma(z)^2} + {\sigma''(z)\over\sigma(z)}
- h_2^{CM} \\
&= \lambda^2 -\lambda h_1\sp{CM}+h_2^{CM}- \wp (z)
\end{array}
\end{equation}
which is a common representation for the $N=2$ Calogero curve (see, for
example formula (\ref{caln2}) below,  with $\lambda =
\lambda' - {\sigma'(z)\over\sigma(z)}$, $x = \wp (z)$ and $h_2^{CM} = -h$).

Returning to the general setting, instead of (\ref{dhph2}) one gets
\be
\sum_{k=0}^N{\tilde\lambda}^{N-k}h_k{\theta_\ast(z+k\epsilon)\over\theta_\ast(z)}\
\rightarrow\ \epsilon^N{\theta_\ast\left(z-{\d\over\d\lambda'}
\right)\over\theta_\ast(z)}\sum_{k=0}^N{\lambda'}^{N-k}h_k^{CM}
+ {\cal O}(\epsilon^{N+1})
\ee
which is the D'Hoker-Phong form of the Calogero curve \cite{DHPhcu} (see
also \cite{vaninsky})
\footnote{In (\ref{dhph2})  we have chosen $\sigma $-functions
for convenience instead of $\theta_\ast$-functions, this differs
slightly from the general formula (\ref{dhph}). However, one
may easily check that this difference is inessential and the result is still
\be
\epsilon^2\ {\theta_\ast\left(z-{\d\over\d\lambda'}\right)\over\theta_\ast(z)}
\left(\lambda'^2 -\lambda'h_1\sp{CM} +h_2^{CM}+2\eta\right) + {\cal
O}(\epsilon^3)= \epsilon^2\left(\lambda^2 -\lambda h_1\sp{CM}+h_2^{CM}- \wp (z)\right) + {\cal
O}(\epsilon^3).
\ee}
\be
\label{dhph}
{\theta_\ast\left(z-{\d\over\d\lambda'}
\right)\over\theta_\ast(z)}\sum_{k=0}^N{\lambda'}^{N-k}h_k^{CM} =
{\theta_\ast\left(z-{\d\over\d\lambda'}
\right)\over\theta_\ast(z)}\ P_N(\lambda' ) = 0.
\ee
The Calogero-Moser-KP ``quasimomentum" (that is the meromorphic differential
with single first order pole and vanishing $\bf A$ periods) is
\be
dp = d\lambda' = d\left(\lambda - {\theta_\ast'(z)\over\theta_\ast(z)}\right)
\ee
and the Seiberg-Witten periods are
\be
a_i = \oint_{A_i}\log\lambda dz, \qquad
a^D_i = \oint_{B_i}\log\lambda dz ,
\ee
in the Ruijsenaars model. The period matrix is, as always
\be
T_{ij} = {\d a^D_i\over\d a_j}, i,j=1,\dots,N.
\ee
\subsection{Theta Functions for the Calogero-Moser Family}

We will next consider several simplifications that arise when the spectral curve
(\ref{laxcu}) covers a 1-dimensional complex torus, or its degenerations, a
cylinder or sphere with two punctures.
This setting contains the elliptic Calogero-Moser and Ruijsenaars-Schneider
models, corresponding to SW theories with adjoint matter, with the degenerations
including the periodic Toda chain corresponding to pure gluodynamics.
In the case of a curve covering a torus the theta functions defined on
the Jacobian of the curve simplify, and these are the expressions we shall
need.

The basic Riemann theta-function function with characteristics is defined by
\begin{equation}
\label{theta}
\Theta\left[\begin{array}{l}{\Bf \epsilon}\\ {\Bf \epsilon'}
            \end{array}\right]
({\bf z}|T)
= \sum_{{\bf n}\in{\Z }^N}
e^{2\pi i\sum_{i=1}^N(n_i+\frac{\epsilon_i}{2})(z_i+\frac{\epsilon_i'}{2}) +
i\pi\sum_{i,j=1}^N(n_i+\frac{\epsilon_i}{2})T_{ij}(n_j+\frac{\epsilon_j}{2}) }
%=\sum_{{\bf n}\in{\Z }^N}
%e^{2\pi i ({\bf n}+\frac{\Bf\epsilon}{2}).({\bf z}+\frac{\Bf\epsilon'}{2})
%+({\bf n}+\frac{\Bf\epsilon}{2}).T.({\bf n}+\frac{\Bf\epsilon}{2})}
\end{equation}
%\be
%\label{theta}
%\Theta ({\bf z}|T) = \sum_{{\bf n}\in{\Z }^N}
%e^{2\pi i\sum_{i=1}^Nn_iz_i + i\pi\sum_{i,j=1}^Nn_iT_{ij}n_j}
%\ee
and we set $\Theta\left[\begin{array}{l}{\Bf 0}\\ {\Bf 0}\end{array}\right]
({\bf z}|T)=\Theta({\bf z}|T)$. By taking $T$ to be the period matrix of
a curve of genus $g=N$ we associate a theta function to the Jacobian of the
curve.
The period matrices of our models satisfy the constraint (\ref{Tconstr})
\be
\label{matrco}
\sum_{i=1}^N T_{ij}\stackrel{\forall j}{=}\ \tau\ \stackrel{\forall
i}{=}\sum_{j=1}^NT_{ij},
\ee
and this allows simplifications.
We will show that
\begin{equation}
\label{thetsim}
\Theta({\bf z}|T)=\sum_{k=0}\sp{N-1}
\theta_{k\over N}(z| N\tau)\Theta_{k}({\bf\hat z}| {\hat T})
\end{equation}
where
$\theta_{k\over N}(z| N\tau)\equiv\theta\left[ \begin{array}{c}{2k\over N}\\ 0
            \end{array}\right](z| N\tau)
$ is a genus $g=1$ or Jacobi theta-function with characteristic,
$\Theta_{k}$ is a genus $N-1$ theta function and $z$, ${\bf\hat z}$ and
${\hat T}$ will be defined shortly.

Let ${\bf e}_N=(1,\ldots 1)$ be  the $N$ vector with all $1$'s, and
from this construct the projection matrix
$P=\frac{1}{N}{\bf e}_N\sp{T}{\bf e}_N$.
Because of (\ref{matrco}) one may write
\be
T=\tau P+ {\tilde T},\qquad {\tilde T}=(1-P)T(1-P).
\ee
Thus $ T_{ij} = {\tau\over N} + {\tilde T}_{ij} $ and
$\sum_{i=1}^N {\tilde T}_{ij}\stackrel{\forall j}{=}\ 0\ \stackrel{\forall
i}{=}\sum_{j=1}^N{\tilde T}_{ij}$.
We may similarly decompose ${\bf z}$,
\be
{\bf z}= {z\over N}\,{\bf e}_N+
{\bf \tilde z}, \qquad {\bf \tilde z}=(1-P){\bf z}.
\ee
Thus $ z_i = {z\over N} + {\tilde z}_i$ and $\sum_{i=1}^Nz_i=z$
or $\sum_{i=1}^N{\tilde z}_i=0$.
In order to express $\Theta({\bf z}|T)$ in the form (\ref{thetsim})
introduce the matrices
\be
M = \left( \begin{array}{ccccr}
  1 & 0 & \ldots &0& -1 \\
  0 & 1 & \ldots &0& -1 \\
  \vdots &  & \ddots && \vdots \\
  0&0&&1&-1\\
  1 & 1 & \ldots&1 & 1
\end{array}\right)
\ \ \ \ \ \ \
M\sp{-1} = \frac{1}{N}\left(
\begin{array}{ccccc}
  N-1 & -1 & \ldots & -1&1 \\
  -1 & N-1 & \ldots & -1&1 \\
  \vdots &  & \ddots && \vdots \\
  -1 & -1 & \ldots & N-1&1\\
   -1&-1&\ldots&-1&1
\end{array}\right)
\ee
and the change of basis
\be
\left( \begin{array}{c} {\bf\hat z}\\ z\end{array}\right)= M{\bf z},
\ee
where ${\bf\hat z}$ is now an $N-1$ vector. Then
\begin{equation}
{\bf n}.{\bf z}={\bf n}\sp{T}M\sp{-1}\, M{\bf z}
=\sum_{j=1}\sp{N-1} (n_j-{k\over N}) {\hat z}_j + \frac{z k}{N}
=({\bf \hat n}-{k\over N}{\bf e}_{N-1}).{\bf\hat z}+ \frac{z k}{N}
\end{equation}
with $k=\sum_{j=1}\sp{N}n_j$. Also
\begin{equation}
{\bf n}\sp{T} T {\bf n}={\bf n}\sp{T}M\sp{-1}\, M(\tau P+\tilde T)M\sp{T}
M\sp{T\,-1}{\bf n}
=\frac{\tau}{N}
k\sp2+{\bf n}\sp{T}M\sp{-1}\, M\tilde TM\sp{T}M\sp{T\,-1}{\bf n}.
\end{equation}
Now
$$
M\tilde TM\sp{T}=
M(1-P)T(1-P)M\sp{T}=\left(\begin{array}{cccc}&&&0\\ &\hat T&&0\\&&&\vdots\\
0&0&\ldots&0\end{array}\right)
$$
where (for $i,j\le N-1$)
$$
{\hat T}_{ij} =T_{ij}-T_{iN}-T_{Nj}+T_{NN}=
{\tilde T}_{ij}-{\tilde T}_{iN}-{\tilde T}_{Nj}+{\tilde T}_{NN}=
{\tilde T}_{ij} + \sum_{k=1}^{N-1}({\tilde T}_{ik} + {\tilde T}_{kj}) +
\sum_{k,l=1}^{N-1}{\tilde T}_{kl}.
$$
Thus
$${\bf n}\sp{T} T {\bf n}=\frac{\tau}{N}k\sp2+
({\bf \hat n}-{k\over N}{\bf e}_{N-1})\sp{T} {\hat T}
({\bf \hat n}-{k\over N}{\bf e}_{N-1}).
$$
In terms of these quantities we see
\be
\label{dec}
\Theta ({\bf z}|T) =
 \sum_{{\bf n}\in{\Z }^N}
e^{2\pi i{\bf n}.{\bf z}+i\pi {\bf n}\sp{T} T {\bf n} } = \\
=
 \sum_{{\bf n}\in{\Z }^N; \sum_{i=1}^N n_i=k}
e^{2\pi i{z k\over N} + i\pi{\tau\over N}k^2 +
2\pi i\sum_{j=1}^{N-1}(n_j-{k\over N}) {\hat z}_j
+ i\pi\sum_{l,m=1}^{N-1}(n_l-{k\over N}){\hat T}_{lm}(n_m-{k\over N})} =
\\
= \sum_{ k\in {\Z }}e^{2\pi i{k\over N}z + i\pi {k^2\over N}\tau}
\sum_{{\bf \hat n}\in\Z^{N-1}}e^{2\pi i({\bf \hat n}-
{k\over N}{\bf e}_{N-1})\hat{\bf z} + i\pi({\bf\hat n}-{k\over N}{\bf e}_{N-1}
)\sp{T} \hat T({\bf\hat n}- {k\over N}{\bf e}_{N-1})}
\ee
By writing $k=Nm + l$ with $m\in{\Z }$ and $l\in {\Z }_N = {\Z }\ {\rm
mod} N$ (i.e. $i=0,1,\dots,N-1$), one has
$$
\Theta ({\bf z}|T)=  \sum_{l\in{\Z }_N}\sum_{m\in{\Z }}
e^{2\pi i\left(m+{l\over N}\right)z + i\pi N\tau\left(m+{l\over N}\right)^2}
\sum_{{\bf \hat n}\in\Z^{N-1}}e^{2\pi i({\bf \hat n}-
{l\over N}{\bf e}_{N-1})\hat{\bf z} + i\pi({\bf\hat n}-{l\over N}{\bf e}_{N-1}
)\sp{T} \hat T({\bf\hat n}- {l\over N}{\bf e}_{N-1})}
$$
That is
\be
\label{decfin}
\Theta ({\bf z}|T) =
 \sum_{l\in{\Z }_N}\theta_{l\over N}(z| N\tau)\Theta_{l}({\bf\hat z}|
{\hat T})
\ee
where $\theta_{l\over N}$ is the genus $g=1$ theta-function introduced
earlier while
\be
\label{Theta}
\Theta_{l}({\bf\hat z}|{\hat T}) =
\sum_{{\bf \hat n}\in\Z^{N-1}}e^{2\pi i({\bf \hat n}-
{l\over N}{\bf e}_{N-1})\hat{\bf z} + i\pi({\bf\hat n}-{l\over N}{\bf e}_{N-1}
)\sp{T} \hat T({\bf\hat n}- {l\over N}{\bf e}_{N-1})}
\equiv
\Theta\left[\begin{array}{c}{2l\over N}{\bf e}_{N-1}\\ 0
            \end{array}\right]
({-\bf\hat z}|\hat T)
\ee
is defined on a $(N-1)$-dimensional complex torus.
(For $N-1>4$ this torus corresponding to $\hat T$, is not necessarily
a Jacobian.)
Our expression (\ref{thetsim}) is equivalent to that obtained in
\cite{BMMM3,M99c} upon observing
\be
\label{thetahat}
\Theta_{k}({\bf\tilde z}|{\tilde T}) =
\sum_{{\bf n}\in{\Z }^N; \sum_{j=1}^N n_j=k}
e^{2\pi i\sum_{j=1}^Nn_j{\tilde z}_j + i\pi\sum_{j,j'=1}^Nn_j{\tilde T}_{jj'}
n_{j'}} 
\\
= \sum_{{\bf m}\in\left({\Z }-{k\over N}\right)^{N-1}}
e^{2\pi i\sum_{j=1}^{N-1}m_j{\hat z}_j + i\pi\sum_{j,j'=1}^{N-1}m_j
{\hat T}_{jj'}m_{j'}} \equiv \Theta_{k}({\bf\hat z}|{\hat T}).
\ee
We have therefore established (\ref{thetsim}) for those models with period
matrices satisfying (\ref{Tconstr}). The coefficients $\Theta_k$ will
appear in our later discussions.

For future reference we note that under the transformation
$\Gamma=\left(\begin{array}{cc}A&B\\ C&D\end{array}\right)\in
Sp(2g,\R)$, where
$A^t D-C^t B = 1_g$, $A^tC= C^t A$ and $B^tD= D^t B$, we have the
arguments of the theta function transforming as
\begin{equation}
\begin{array}{rcl}
 T  & \rightarrow &  T ^{\Gamma} = (A  T  + B) (C  T  + D)^{-1} ~, \\
z & \rightarrow & z^{\Gamma} = [ (C T  + D)^{-1}]^t z ~.
\end{array}
\end{equation}
Further, the characteristics transform as
\begin{equation}
\begin{array}{rcl}
\epsilon &\rightarrow &
\epsilon ^{\Gamma}= D\epsilon-C\epsilon' +{1 \over 2} {\rm
diag}(CD^t) ~, \\
\epsilon'  &\rightarrow & \epsilon ^{\prime\,\Gamma}=
-B\epsilon+A\epsilon' +{1 \over 2} {\rm
diag}(AB^t) ~.
\end{array}
\end{equation}
For $\Gamma=\left(\begin{array}{cc}M&0\\ 0&M\sp{-1\,T}\end{array}\right)$
we see
$$
T ^{\Gamma} =M TM\sp{T}=
\left(\begin{array}{cccc}&&&0\\ &\hat T&&0\\&&&\vdots\\
0&0&\ldots&N\tau\end{array}\right),\quad
z^{\Gamma} =\left( \begin{array}{c} {\bf\hat z}\\ z\end{array}\right),
$$
and our previous discussion has simply used this transformation to put
the theta function into a canonical form. Further the particular case
\be
\label{thetainv}
\Theta({\bf z}|T)=\zeta \left(\det T\right)\sp{-1/2}
e\sp{-i\pi \bf z\sp{T}T\sp{-1}\bf z}\,
\Theta({T\sp{-1}\bf z}|-T\sp{-1}).
\ee
(for some $\zeta\sp8=1$) leads to
\begin{equation}
\begin{array}{rl}
\Theta({\bf z}|T)&=\sum_{k=0}\sp{N-1}
\Theta\left[\begin{array}{c}{k\over N}\\ 0\end{array}\right]
(z|N\tau)
\Theta\left[\begin{array}{c}{k\over N}{\bf e}_{N-1}\\ 0
            \end{array}\right]
({-\bf\hat z}|\hat T)\\
&=\zeta \left(\det \hat T\right)\sp{-1/2}\left(N\tau\right)\sp{-1/2}
e\sp{-i\pi{\bf\hat z} \hat T\sp{-1}{\bf\hat z} -i\pi {z\sp2 \over {N\tau}}}
\sum_{k=0}\sp{N-1}\,
\Theta({z\over N\tau}+{k\over N}|{-1\over N\tau})\,
\Theta(\hat T\sp{-1}{\bf\hat z} -{k\over N}{\bf e}_{N-1}|-\hat T\sp{-1}).
\end{array}
\end{equation}

\subsection{Elliptic solutions}
We conclude this section by examining some explicit solutions to the
integrable systems described earlier.
The simplest non degenerate periodic solutions arise in integrable
systems with only two interacting particles. Because the
center of mass decouples in the Calogero-Moser families we have been
considering there is effectively one degree of freedom in this case.
We will consider the ``periodic" (sine-Gordon) Toda chain and Calogero-Moser models.

For the example of the ``periodic" Toda chain with two
particles an explicit solution is a simple
consequence of the addition formula for the (Weierstrass) elliptic functions
\be
\wp (\mu+\xi ) + \wp (\mu-\xi ) -2\wp (\mu) = - {\d^2\over\d \mu^2}\log\left(
\wp(\mu) - \wp(\xi )\right).
\label{add}
\ee
Indeed, consider
\be
e^q = A{\sigma (\mu+\omega)\over\sigma(\mu)\sigma(\omega)}e^{-\eta \mu},
\ee
where $\mu=Ut$, $\omega$ is any of the (half-) periods (
$\wp (\mu+2\omega) = \wp (\mu)$ and $\eta = \zeta(\omega)$) and the constants
$A$ and $U$ have yet to be determined. Then
\begin{equation}
\begin{array}{rl}
e^{2q}& = A^2{\sigma^2 (\mu+\omega)\over\sigma^2(\mu)\sigma^2(\omega)}e^{-2\eta
\mu} =
- A^2{\sigma (\mu+\omega)\sigma (\mu-\omega)\over\sigma^2(\mu)\sigma^2(\omega)}
=
A^2\left(\wp (\mu) - \wp(\omega)\right),
\\
e^{-2q} &= A^{-2}{1\over\wp (\mu) - \wp(\omega)} = {1\over A^2 H\sp2 }
\left(\wp (\mu+\omega ) - \wp(\omega)\right).
\end{array}
\end{equation}
Here $ H\sp2 = (e-e_+)(e-e_-)$, with $e=\wp(\omega )$ and
$e_\pm=\wp(\omega_\pm )$,
where $\omega_\pm$ are the remaining half-periods: $\omega+
\omega_++\omega_-=0$ modulo the lattice. If one puts $A^4 H\sp2  = 1$ then
\begin{equation}
\begin{array}{rl}
e^{2q} - e^{-2q} &= {1\over{ H}}\left(\wp (\mu+\omega ) -
\wp(\mu )\right) =
{1\over{2 H}}\left(\wp (\mu+\omega ) + \wp (\mu-\omega ) -
2\wp(\mu )\right) =
- {1\over{2 H}}{\d^2\over\d \mu^2}\log\left(\wp(\mu) -
\wp(\omega )\right)
\\ &= -{1\over{2 H}}{\d^2\over\d \mu^2}\log{\sigma (\mu+\omega)
\sigma(\omega-\mu)\over
\sigma^2(\mu)\sigma^2(\omega)} = -{1\over{2 H}}{\d^2\over\d \mu^2}
\log e^{2q} =
- {1\over{ H}}{\d^2 q\over\d \mu^2} = - {1\over{H U^2}}
{d^2 q\over dt^2}.
\label{eqmo}
\end{array}
\end{equation}
After introducing $\Lambda$ via
\be
U = {\Lambda\over i \sqrt{H}}
\ee
formula (\ref{eqmo}) acquires the ``canonical" form
\be
\label{eqmo1}
{d^2 q\over dt^2} = \Lambda^2\left( e^{2q} - e^{-2q}\right)
\ee
with $\Lambda^2$ playing the role of a coupling constant. We may rescale
$\Lambda\to1$, in which case these are the equations of motion for
\be
\label{hamsu2}
h = p^2 + e^{2q} + e^{-2q} \equiv p^2 + 2\cosh 2q
\ee
which is the periodic Toda Hamiltonian  (\ref{Toda2}) in the centre of
mass frame. (Here $q=q_1=-q_2$.) The Lax operator for this example is
\be
\label{laxsu2}
{\cal L}(w) = \left(
\begin{array}{cc}
  p & e^{-q} + {1\over w}e^q \\
  e^{-q} + we^q & -p
\end{array}\right)
\ee
and the corresponding spectral curve in this case
\be
\label{scsu2}
w + {1\over w} = \lambda^2 - h.
\ee
The scale parameter $\Lambda$ may be restored in the above by
setting $p\to p/\Lambda$, $\lambda\to\lambda/\Lambda$ and $w\to w/\Lambda\sp2$.
Finally the perturbative limit leading to (\ref{CM2pert}) is obtained by
setting $p\to p/\Lambda$, $\lambda\to\lambda/\Lambda$, $w\to w/\Lambda\sp2$
and also $q\to q-\log\Lambda$ and taking the $\Lambda\to0$ limit.
Another convenient representation for the spectral curve (\ref{scsu2}) is
the elliptic parameterization
\be
w^3 + hw^2 + w = \lambda^2w^2,
\ee
where
\be
\label{ellparaa}
\begin{array}{rl}
w &= x - e = \wp(z) - e = \wp(z) - \wp(\omega) = - {\sigma
(z+\omega)\sigma(z-\omega)\over\sigma^2(z)\sigma^2(\omega)} \\
&= \left({\sigma(z+\omega)\over\sigma(z)\sigma(\omega)}
e^{-\zeta(\omega)z}\right)^2
 = \left({\sigma(z-\omega)\over\sigma(z)\sigma(\omega)}
e^{\zeta(\omega)z}\right)^2
\end{array}
\label{ellpar}
\ee
and
\be
\lambda = \2{\wp'(z)\over\wp (z) -e} .
\ee
In this form $h=e$, $e_+ e_-=1$ and $g_2=4(h\sp2 -1)$, $g_3=4 h$.
Also $dz = 2{dw\over\lambda w}$.
Observe that $\lambda\to\infty$ as $w\to0,\infty$ (equivalently
$z\to\omega,0$ or $x\to e,\infty$).
Let us denote $P_+$ to be the point $\lambda\to\infty$, $w\to\infty$
and $P_-$ to be the point $\lambda\to\infty$, $w\to0$.
Further let $P(1)$ be the point $\lambda=p$, $w=-e\sp{-2q}$
and $P(0)$ the point $\lambda=-p$, $w=-e\sp{2q}$.

The solution we have just obtained may be derived from
the Baker-Akhiezer function for this problem.
The auxiliary linear problem ${\cal L}\Psi = \lambda\Psi$ has solution
\be
\label{basu2}
\Psi
= \left(
\begin{array}{c}
  \psi_0 \\
   \psi_1
\end{array}\right) = \left(
\begin{array}{c}
   1 \\
   {we^q + e^{-q}\over \lambda + p}
\end{array}\right)
\ee
while the ``conjugate" equation ${\hat\Psi}{\cal L} =\lambda{\hat\Psi}$ has
solution
\be
\label{conjug}
{\hat\Psi} =({\hat\psi}_0, {\hat\psi}_1)=
 \left( {we^q + e^{-q}\over \lambda - p }, 1\right).
\ee
Consider the expression
$$
{\psi_1\over\psi_0} = {we^q + e^{-q}\over \lambda + p}.
$$
From our earlier remarks this vanishes at $P_-$ and because $w\sim \lambda\sp2$
as $\lambda\to\infty$ it has a pole at $P_+$.
Also there is a further zero at $P(1)$ and a pole at $P(0)$, and so we
have the divisor of $\psi_k$
\be
(\psi_k) = P(k) - P(0) - kP_+ + kP_- .
\ee
Now a generic Baker-Akhiezer function should have the same number of poles and
zeros.  Introducing
\be
\psi_k = w^{k/2}\tilde\psi_k
\ee
gives
\be
{\psi_1\over\psi_0}= w^{1/2}
{\tilde\psi_1\over\tilde\psi_0}
\ee
and
\be
\label{dtpsi}
(\tilde\psi_i) = P(i) - P(0).
\ee
In particular we may express the the Baker-Akhiezer function in terms of
the elliptic parameterisation (\ref{ellparaa}) as
\be
\tilde\psi_i ={\rm{const.}}\ \frac
{\theta\left(z-z(P((i)))\right)}{\theta\left(z-z(P(0))\right)}.
\ee

We conclude this section with the simplest example of the
elliptic Calogero-Moser model, that corresponding to two particles.
In the centre of mass frame $h_1\sp{CM}=p_1+p_2=0$ eqn.~(\ref{CM2}) turns into
\be\label{caln2}
\lambda ^2 + h - x = 0,
\ee
where $h=-h_2\sp{CM}=-p_1p_2+\wp(q_{12})$, $x=\wp(z)$ and we have
removed the $m^2$ dependence in (\ref{CM2}) using the scaling properties
(see Appendix A) of the $\wp$ function.
This equation says that to any value of $x$ one
associates two points of $\Sigma^{CM}$
\be
\lambda = \pm\sqrt{h - x},
\ee
i.e. it describes $\Sigma^{CM}$ as a
double covering of an elliptic curve ramified at the points $x = h$ and
$x = \infty$. In fact, $x = h$ corresponds to a
{\em pair} of points distinguished by the sign of $y$, but $x = \infty$
is one of the branch points, and so the {\em
two} cuts between $x = h$ and $x=\infty$ on each sheet become effectively
a single cut between $(h, +)$ and $( h, -)$.
The spectral curve $\Sigma^{CM}$ may therefore be considered
as two tori glued along one cut, and so $\Sigma^{CM}_{N=2}$ has
genus 2. This is a hyperelliptic curve (for $N = 2$ only!).
The two holomorphic 1-differentials on $\Sigma^{CM}$ ($g = N = 2$) can be
chosen to be
\be\label{holn2}
dv_+ = dz = \frac{dx}{2y} = \frac{\lambda d\lambda}{y}
\ \ \ \ \ \ \
dv_- = {dz\over\lambda} = \frac{dx}{2y\lambda}=\frac{d\lambda }{y}
\ee
so that
\be
dS = 2\lambda dz = \lambda{dx\over y} =
\frac{dx}{y}\sqrt{h - x},
\ee
and
\be
- \frac{\partial dS}{\partial h} \cong \frac{dx}{2y\lambda} = dv_- .
\label{CdSd}
\ee
The fact that only one of the two holomorphic 1-differentials (\ref{holn2})
appears on the right hand side of (\ref{CdSd}) is related to their different
parities with respect to the ${\Z}_2\otimes {\Z}_2$ symmetry of $\Sigma^{CM}$:
$y \rightarrow -y$, $\lambda \rightarrow -\lambda$ and
$dv_{\pm}\rightarrow \pm dv_{\pm}$.
Since $dS$ has  a positive parity, its
integrals along two of the four elementary non-contractable cycles
on $\Sigma^{CM}$ automatically vanish, leaving only two
non-vanishing quantities $a$ and $a\sp{D}$, as necessary for
the SW interpretation.
Moreover, the two remaining nonzero periods can be defined
in terms of a ``reduced" curve of genus $g=1$
\be
Y^2 =  (y\lambda)^2 =4 \left(h - x\right)
 \prod_{a=1}^3 (x - e_a),
\label{calN2red}
\ee
equipped with $dS = \left(h - x\right)\frac{dx}{Y}$. This curve arises
when directly integrating the equations of
motion since after decoupling the free motion of the center of mass we again
have a dynamical system with only one degree of freedom.
From the conserved energy $h = p^2 + \wp (q)$ we obtain
\be
\label{intred}
t = \int {dq\over\sqrt{h - \wp (q )}} = \int {dx\over
\sqrt{4(x-e_1)(x-e_2)(x-e_3)(h - x)}}
\ee
which is exactly the Abel map of the reduced curve (\ref{calN2red}).

\section{The weak coupling limit}

In this section we are going to study the weak coupling limit in some detail.
First we consider the $\Lambda\to 0$ limit of the periodic Toda chain,
demonstrating that this leads exactly to the open Toda chain or Toda molecule.
Here we explicitly derive formulae for the theta-functions (tau-functions) in
this limit and show that they appear to be a finite-dimensional analogue of
those appearing in matrix models. We discuss their relation with duality in
integrable systems and the commutativity of theta-functions \cite{M99c}.
We then turn briefly to the corresponding properties of the weak coupling
limit of Calogero-Moser models.

\subsection{From the periodic Toda chain to the Toda molecule}
The perturbative limit corresponds to $\tau\to +i\infty$ and
$\Lambda\to 0$. The effect of this will be to simplify the theta function
solutions describing the general finite gap situation.

First, in the limit $\tau\to +i\infty$, the Jacobi theta-functions in
(\ref{thetsim}) turn into exponentials and one finds that
\be\label{dectoda}
\Theta ({\bf z}|T) = \sum_{k\in{\Z }_N}
e^{2\pi i{k\over N}z} \Theta_{k}.
\ee
We may already deduce an interesting consequence:
the ratios of the coefficients $\Theta_k\equiv\Theta_{k}({\bf\hat z}|{\hat T })
\equiv\Theta_{k}({\bf\tilde z}|{\tilde T})$ Poisson commute,
\be
\label{poicom}
\left\{ {\Theta_i\over\Theta_j}, {\Theta_{i'}\over\Theta_{j'}}\right\} = 0,
\qquad
\forall\ i,j,i',j'.
\ee
The Poisson bracket here is that corresponding to the symplectic form
\be
\label{sympfor}
\Omega^{Toda} = \sum_{i=1}^{N-1}d{\hat z}_i\wedge da_i =
\sum_{i=1}^{N-1}d{\tilde q}_i\wedge d{\tilde p}_i,
\ee
where ${\tilde q}_i$ and ${\tilde p}_i$ are co-ordinates and momenta of
the Toda chain
and we are working in the centre of mass frame ${\tilde q}_i=q_i-q_{\rm {CM}}$.
The Poisson commutativity
of the ratios (\ref{poicom}) follows from the solution of the periodic
Toda chain \cite{DaTa,KriTo}
\be
\label{todasol}
e^{{\tilde q}_i} = {\Theta_i\over\Theta_{i-1}},
\qquad
{\Theta_i\over\Theta_j} = \prod_{k=j+1}^i e^{{\tilde q}_k}.
\ee
(Here  $\Theta_0=\Theta_N$ and the $\Theta_k$ can be thought of as the
Toda chain tau-functions depending on the discrete time $k$, the
number of the particle. Observe $\sum_i {\tilde q}_i=0$ here.)
Now because the coordinates $\tilde q_i$ obviously Poisson commute,
$\{ \tilde q_i,\tilde q_j\} = 0$, we deduce (\ref{poicom}).
This expression gives a precise formulation of the old expectation that
the Toda chain tau-functions Poisson commute with each other.
We will return to this point shortly when we discuss duality.
Henceforth we shall drop the tilde from the coordinates and momenta for
simplicity.

The second set of simplifications arise because the spectral curve
(\ref{todacu})  degenerates into a rational curve (\ref{polyn})
\be
\label{optocu}
w = P_N(\lambda) = \prod_{i=1}^N(\lambda-a_i)
\ee
($\sum a_i = 0$) in the perturbative limit $\Lambda\to 0$.
Then upon making the natural choice (for the cylinder) $z=\log w$
\be a_i = \oint_{A_i}\lambda{dw\over w} \ee
and we have the basis of (normalised) holomorphic differentials
\be
\label{degholdif}
2\pi id\omega_i = {d\lambda\over\lambda-a_i}.
\ee
The perturbative period matrix ${\tilde T}_{ij} = {\d a^D_i\over\d a_j}$
(where $a^D_i = \oint_{B_i}\lambda{dw\over w}$) is given by
\be
\label{tpert}
-i\pi{\tilde T}_{ij} \equiv -i\pi{\tilde T}_{ij}^{\rm pert} =
\delta_{ij}\sum_{l\neq i}\log\frac{ |a_{il}|}{\Lambda}
 - (1-\delta_{ij})\log\frac{ |a_{ij}|}{\Lambda}.
\ee
We shall substitute this into (\ref{dectoda}) and take the $\Lambda\to0$ limit,
but first we should be more careful with the various appearances of
$\Lambda$. We have already mentioned the scalings $h_k\to h_k/\mu^k$,
$w\to w/\mu^N$ along with $\lambda\to \lambda/\mu$. Comparison of
(\ref{dectoda}) and  the relation between $w\sim e^{z+i\pi\tau}$ and
$z$ shows the first two of
these scalings to be achieved by $2\pi i z\to 2\pi i z-N\log\mu$.
Now the double scaling limit (where $\Lambda=m\, e^{i\pi\tau/N}$) is
achieved by shifting $2\pi i z\to 2\pi i z-N^2\log\Lambda$. (We shall justify
this a little later.) Upon noting the $\tau$ dependence of the genus one
theta function we see $\Theta_k\to\Lambda^{k^2}\Theta_k$ and
overall we must construct
\be
\label{thetaop}
\bar\Theta_k \equiv
{\lim_{\Lambda\to0}}\ \Lambda^{k^2-N k}\Theta_k = \\
=\sum_{\sum_{i=1}^Nn_i=k;\ \sum_{i<j} (n_i-n_j)^2=k(N-k)}
e^{2\pi i\sum_{j=1}^Nn_j{\tilde z}_j + i\pi\sum_{i,j=1}^Nn_in_j{
( (1-\delta_{ij})\log|a_{ij}|}-\delta_{ij}\sum_{l\neq i}\log|a_{il}|) } =
\\
= \sum_{\sum_{i=1}^Nn_i=k;\ \sum_{i<j} (n_i-n_j)^2=k(N-k)}
e^{2\pi i\sum_{j=1}^Nn_j{\tilde z}_j - i\pi\sum_{i<j=1}^N(n_i-n_j)^2{
\log|a_{ij}|} } .
\ee
The quadratic constraint here appears when the $\Lambda$ dependence of $\tilde
T_{ij}$ is taken into account.
Now  the conditions $\sum n_i = k$ and $\sum_{i<j} (n_i-n_j)^2=k(N-k)$
can only be satisfied for each $n_i\in\{0,1\}$, with $k$ of these nonzero.
Thus we may write $\bar\Theta_k$ as
\be
\label{bartheta}
\bar\Theta_k = \sum_{|I|=k}\prod_{i\in I}e^{2\pi i\tilde z_i}
\prod_{j\in{\bar I}} \frac{1}{|a_{ij}|}
\ee
for some set of indices $I=\{ i_1,\dots,i_k\}$,
and  ${\bar I}$ is the set of indices complementary to $I$
($n_i=1$ if $i\in I$ and $n_i=0$ otherwise).
These are expressions for the tau-functions of the {\em open} Toda chain or
the {\em Toda molecule}.

Let us consider the $N=2$ version of these formulae. For this
case of one degree of freedom we have the hamiltonian (\ref{CM2pert})
$H = p^2 + e^{2q}$.
Taking for this example the period matrix
\be
-i\pi\|{\tilde T}_{ij}\| = \left(
\begin{array}{cc}
  \log{\frac{|a_1-a_2|}{\Lambda}} & -\log{\frac{|a_1-a_2|}{\Lambda}}  \\
  -\log{\frac{|a_1-a_2|}{\Lambda}}  & \log{\frac{|a_1-a_2|}{\Lambda}}
\end{array}\right) \equiv \left(
\begin{array}{cc}
  \log \frac{a}{\Lambda} & -\log \frac{a}{\Lambda}  \\
  -\log \frac{a}{\Lambda}  & \log \frac{a}{\Lambda}
\end{array}\right)
\ee
formula (\ref{thetaop}) gives in this case
$\bar\Theta_0 = 1$ and
\be
\bar\Theta_1 = \sum_{i+j=1; (i-j)^2=1}e^{(i-j)z-i\pi(i-j)^2{\log a}},
\ee
where we have substituted $2\pi i\tilde z_1\to z$, $2\pi i\tilde z_2\to -z$.
Thus
\be
\label{theta1}
\bar\Theta_1 = \sum_{i+j=1; (i-j)^2=1}e^{-(i-j)^2\log a + (i-j)z} =
{1\over a}(e^z + e^{-z}).
\ee
Now $e\sp{2q}=e\sp{q_2-q_1}=(\bar\Theta_2/\bar\Theta_1)\sp2=1/\bar\Theta_1\sp2$
and so we again arrive at the explicit solution to the equation of
motion\footnote{It is interesting to point out that the equation of motion for
the co-ordinate $Q=a=\sqrt{h}$ in the dual system $H={\cosh P\over Q}$
coincides with the equation of motion for the rational Calogero model, $\ddot{Q}
={1\over Q^3}$. Similar dualities are known for the Calogero models
and their ``relativistic" counterparts.}
for the hamiltonian $h = p^2 + e^{2q}$
\be
\label{liusol}
e^{q} = {\sqrt{h}\over\cosh z}.
\ee
In the $N=3$ case eq.~(\ref{bartheta}) gives
\be
\bar\Theta_1 = e^{2\pi i\tilde z_1}{1\over |a_{12}a_{13}|} +
e^{2\pi i\tilde z_2}
{1\over |a_{12}a_{23}|} + e^{2\pi i\tilde z_3}{1\over |a_{13}a_{23}|}
\ee
and one further nontrivial $\bar\Theta$ function
\be
\bar\Theta_2 = e^{2\pi i(\tilde z_1+\tilde z_2)}{1\over |a_{13}a_{23}|} +
e^{2\pi i(\tilde z_1+\tilde z_3)}
{1\over |a_{12}a_{23}|} + e^{2\pi i(\tilde z_2+\tilde z_3)}
{1\over |a_{12}a_{13}|} = \\ =
e^{-2\pi i\tilde z_1}{1\over |a_{12}a_{13}|} +
e^{-2\pi i\tilde z_2}{1\over |a_{12}a_{23}|} + e
^{-2\pi i\tilde z_3}{1\over |a_{13}a_{23}|}.
\ee
There is a convenient determinantal form for formula (\ref{bartheta}).
In general one has that
\be
\label{moma1}
\bar \Theta_k =\det_{k\times k}K_{n+m}\Big|_{n,m=1,\dots,k}
\ee
where the ``moment matrix" $K_{nm}=K_{n+m}$ is defined as an average
with respect to $\bar \Theta_1$, i.e.
\be
\label{moma}
K_n = \langle a^n\rangle_1 = \sum_{i=1}^N e^{Z_i}a_i^{n-2}, \qquad
e^{Z_i} \equiv e^{2\pi i\tilde z_i}{\prod_{j\neq i}}{1\over |a_{ij}|},\qquad
\bar \Theta_1 = \sum_{i=1}^N e^{Z_i}.
\ee
The proof is similar to that in matrix models \cite{KMMOZ}. Indeed,
\begin{equation}
\label{mamo}
\begin{array}{rl}
\bar \Theta_k &=
{\sum}_{|I|=k} \prod_{i\in I}e^{2\pi i\tilde z_i}{\prod_{j\in{\bar I}}}
{1\over |a_{ij}|} =
{\sum}_{|I|=k} \prod_{i\in I}e^{2\pi i\tilde z_i}\left({\prod_{j\in{\bar I}}}
{1\over |a_{ij}|}{\prod_{j\in I\setminus\{i\}}}{1\over |a_{ij}|}
                 {\prod_{j\in I\setminus\{i\}}}|a_{ij}|\right)  \\
&={\sum}_{|I|=k}\prod_{i\in I}\left(e^{Z_i}
{\prod_{j\in I\setminus\{i\}}}|a_{ij}|\right) =
{\sum}_{|I|=k}\prod_{i\in I}e^{Z_i}{\prod_{i\ne j\in I}}|a_{ij}| =
{\sum}_{|I|=k}\prod_{i\in I}e^{Z_i}{\prod_{i<j;\ i,j\in I}}a_{ij}^2
\end{array}
\end{equation}
looks like a ``discrete" analogue of the tau-function of the ``forced"
Toda chain hierarchy which plays a central role in matrix models \cite{KMMOZ}.
Now\footnote{This is a particular case of the Cauchy-Binet formula for
the $k\times N$  ($k\leq N$) rectangular matrices $A_{ni}$
and $B_{im}$ ($i=1,\dots,N$, $n,m=1,\dots,k$)
\be
\left.\det_{k\times k}\left(\sum_{i=1}^N A_{ni}B_{im}\right)
\right|_{n,m=1,\dots,k} =
\sum_{i_1<\dots <i_k}
\left.\det_{k\times k} A_{ni}\right|_{i\in I;\ n=1,\dots,k}
\left.\det_{k\times k} B_{im}\right|_{i\in I;\ n=1,\dots,k}.
\ee}
for {\em any} coefficients $C_i$
\begin{equation}
\begin{array}{l}
{\sum}_{|I|=k}\prod_{i\in I}C_i\,{\prod_{i<j;\ i,j\in I}}a_{ij}^2\equiv
\sum_{I:\ i_1<\dots <i_k}C_{i_1}\dots C_{i_k}\prod_{i_n<i_m}\left(a_{i_n}-
a_{i_m}\right)^2
\\
\qquad= \sum_{i_1<\dots <i_k}C_{i_1}\dots C_{i_k}
\left.\det_{k\times k} a_i^{n-1}\right|_{i\in I;\ n=1,\dots,k}
\left.\det_{k\times k} a_i^{m-1}\right|_{i\in I;\ m=1,\dots,k}
\\
\qquad= \left.\det_{k\times k}\left(
\sum_{i=1}^NC_ia_i^{n+m-2}\right)\right|_{n,m=1,\dots,k}.
\end{array}
\end{equation}
Substituting $C_i=e^{Z_i}$, one arrives at eqs.~(\ref{moma1}), (\ref{moma})
\be
\label{mamodet}
\bar\Theta_k = \sum_{I:\ i_1<\dots <i_k}e^{Z_{i_1}+\dots Z_{i_k}}
\prod_{i_n<i_m}\left(a_{i_n}-
a_{i_m}\right)^2 =
\left.\det_{k\times k}\left(
\sum_{i=1}^Ne^{Z_i}a_i^{n+m-2}\right)\right|_{n,m=1,\dots,k} =
\det_{k\times k}K_{n+m}.
\ee
These formulae in fact coincide with the solution of the $N$-particle {\em
open} Toda chain problem in terms of the action-angle variables discussed in
\cite{Toda}.

Let us relate these formulae to the  Baker-Akhiezer function.
First, we have defined the angle variables as
co-ordinates of the Jacobian of genus $g=N$ curve $\Sigma^{N}$ by
(\ref{angles}). For the holomorphic differentials (\ref{degholdif})
we have
\be
\label{zlambda}
2\pi iz_i = \sum_{k=1}^N\int_{P_0}^{P_k}{d\lambda\over\lambda-a_i} =
\sum_{n=0}^{\infty} a_i^n\sum_{k=1}^N\int_{P_0}^{P_k}
{d\lambda\over\lambda^{n+1}}\
\stackreb{\lambda_k\to\lambda(P_0)=\infty}{\sim}\
N\log\lambda + \sum_{n=1}^{\infty}a_i^nT_n .
\ee
Here $\lambda_k\equiv\lambda(P_k)$ and
\be
\label{miwa}
T_n = - {1\over n}\sum_{k=1}^N\lambda_k^{-n}
\ee
is the so called Miwa parameterization of the ``canonical" Toda times.
The logarithmically divergent first term here is absorbed
into the renormalisation $z_i\to z_i - {N\over 2\pi i}\log\Lambda$. Summing
over $i$ yields $z\to z-{N^2\over 2\pi i}\log\Lambda$ and the
double scaling limit we have
already mentioned. The remaining ``finite" part of the right hand side here
describes the Toda chain dynamics with respect to the various
higher times $t_n$, $n\ge1$
\be
2\pi iz_i = a_it + \sum_{n>1}a_i^nt_n + z_i^{(0)}.
\ee
The ordinary time $t_1=t$ here corresponds to the evolution with respect to the
Hamiltonian quadratic in the  momenta (or quadratic in
canonical action variables)
\be
2\pi iz_i = a_it + z_i^{(0)}.
\ee
The  Baker-Akhiezer functions on the degenerate spectral curve
(\ref{optocu}) may now be defined by the following analytic requirements:
%\begin{itemize} \item
It is the set of functions $\psi_k=\psi_k(\lambda )$, $k=0,\dots,N-1$
which have
exactly $k$ zeroes on the rational curve (\ref{optocu}) and a single pole of
order $k$ at $\lambda=\infty$. For $k\geq N$ they can be defined by
$\psi_{k+N}=w\psi_k$.
%\end{itemize}
This means each $\psi_k(\lambda )$, $k=0,\dots,N-1$ may be represented
by a polynomial with $k$ (finite) zeroes.
Thus $\psi_k$ can be constructed as a linear combination of
$\prod_{i_1<\dots <i_k}(\lambda-a_{i_l}) = \prod_{j\in I}(\lambda-a_j)$.
In fact one has
\begin{equation}
\label{ba}
\psi_k(\lambda) = \lambda^k{\bar\Theta_k\left( z_i -
\frac{1}{2i\pi} \sum_{n\ge1}{a_i^n\over
n\lambda^n} \right)\over\bar\Theta_k\left( z_i \right)} =
{{\sum}_{|I|=k}
 \prod_{i\in I}e^{2\pi iz_i}(\lambda - a_i){\prod_{j\in{\bar I}}}
{1\over |a_{ij}|}\over
{\sum}_{|I|=k} \prod_{i\in I}e^{2\pi iz_i}{\prod_{j\in{\bar I}}}
{1\over |a_{ij}|}}.
\end{equation}
Thus, for example
\be
\psi_1(\lambda) = \lambda{\bar\Theta_1\left( z_i - \frac{1}{2i\pi}\sum_{n\ge1}
{a_i^n\over n\lambda^n} \right)\over\bar\Theta_1\left( z_i \right)} =
{\sum_{i=1}^N (\lambda - a_i)e^{2\pi iz_i}{\prod_{j\ne i}}{1\over |a_{ij}|
}\over \sum_{i=1}^N e^{2\pi iz_i}{\prod_{j\ne i}}{1\over |a_{ij}|}} =
\\
= \lambda -
{\sum_{i=1}^N a_ie^{2\pi iz_i}{\prod_{j\ne i}}{1\over |a_{ij}|}\over
\sum_{i=1}^N e^{2\pi iz_i}{\prod_{j\ne i}}{1\over |a_{ij}|}}\equiv
\lambda- \langle a\rangle
\label{psi1n2}
\ee
with $\langle a\rangle\equiv\langle a^3\rangle_1/\langle a^2\rangle_1$.
We note that these Baker-Akhiezer functions satisfy
$$
\sum_{i=1}\sp{N} e\sp{Z_i}\psi_k(a_i)=0
$$
which is a consequence of the identity
$$
\sum_{i=1}\sp{N}\prod_{j\ne i}\frac{1}{a_i-a_j}=0.
$$
(This may be established using Lagrange interpolation:
$1=\sum_{i=1}\sp{N}\prod_{j\ne i}\frac{x-a_j}{a_i-a_j}$.)
One may also easily write the equations of motion for the zeroes of
the Baker-Akhiezer functions $\psi_k(\gamma_i) = 0$:
\footnote{They can be considered, for example, as a degeneration of corresponding
equations for the zeroes of the Baker-Akhiezer function of periodic Toda chain
\be
\label{dueper}
{\d\gamma_i\over\d t} = {\sqrt{P_N^2(\gamma_i)-4\Lambda^{2N}}\over
\prod_{j\neq i}(\gamma_i-\gamma_j)},
\qquad
{\d\gamma_i\over\d t_n} = {\gamma_i^{n-1}\sqrt{P_N^2(\gamma_i)-4\Lambda^{2N}}
\over\prod_{j\neq i}(\gamma_i-\gamma_j)}.
\ee
}
\be
\label{due}
{\d\gamma_i\over\d t} = {\prod_{j=1}^N(\gamma_i-a_j)\over\prod_{j\neq
i}(\gamma_i-\gamma_j)},
\qquad
{\d\gamma_i\over\d t_n} = {\gamma_i^{n-1}\prod_{j=1}^N(\gamma_i-a_j)\over
\prod_{j\neq i}(\gamma_i-\gamma_j)}.
\ee
Finally let us note that the conserved Hamiltonians are straightforwardly
given in terms of the minors of the Lax matrix.
The Lax representation gives
\be
\det(\lambda - {\cal L}) =
\prod_{i=1}^N(\lambda-a_i) = \sum_{k=0}^N\lambda^{N-k}(-1)^kh_k
\ee
with $h_0\equiv 1$, and
\be
h_k = \sum_{i_1< \dots <i_k}a_{i_1}\dots a_{i_k}
\ \ \ \ \ \ \ k=1,\dots,N.
\ee
On the other hand, since (for {\em any} matrix ${\cal L}$)
\be
\label{fr11}
\det\,(\lambda - {\cal L}) = \\ =
\lambda^N - \lambda^{N-1}\sum_i {\cal L}_{ii}+
\lambda^{N-2}\sum_{i<j}\det\left(\begin{array}{cc}{\cal L}_{ii}&{\cal L}_{ij}
\\ {\cal L}_{ji}&{\cal L}_{jj}\end{array}\right)-
 \lambda^{N-3}\sum_{i<j<k}\det
\left(\begin{array}{ccc}{\cal L}_{ii}&{\cal L}_{ij}&{\cal L}_{ik}\\
{\cal L}_{ji}&{\cal L}_{jj}&{\cal L}_{jk}\\ {\cal L}_{ki}&{\cal L}_{kj}
&{\cal L}_{kk}\end{array}\right)+\ldots =
\\
= \lambda^N - \lambda^{N-1}\Tr{\cal L} + \sum_{k=2}^N(-1)^k\lambda^{N-k}
\sum_{i_1< \dots <i_k}\det\left(\begin{array}{cccc}
  {\cal L}_{i_1i_1} & {\cal L}_{i_1i_2} & \dots & {\cal L}_{i_1i_k} \\
  {\cal L}_{i_2i_1} & {\cal L}_{i_2i_2} & \dots & {\cal L}_{i_2i_k} \\
  \vdots &  &\ddots  & \vdots \\
  {\cal L}_{i_ki_1} & {\cal L}_{i_ki_2} & \dots & {\cal L}_{i_ki_k}
\end{array}\right)
\ee
we have that
\be
h_k = \sum_{i_1< \dots <i_k}a_{i_1}\dots a_{i_k} =
\sum_{i_1< \dots <i_k}\det\left(\begin{array}{cccc}
  {\cal L}_{i_1i_1} & {\cal L}_{i_1i_2} & \dots & {\cal L}_{i_1i_k} \\
  {\cal L}_{i_2i_1} & {\cal L}_{i_2i_2} & \dots & {\cal L}_{i_2i_k} \\
  \vdots &  &\ddots  & \vdots \\
  {\cal L}_{i_ki_1} & {\cal L}_{i_ki_2} & \dots & {\cal L}_{i_ki_k}
\end{array}\right).
\ee

\subsection{Theta-functions for the Calogero-Moser models and their
 Poisson Commutativity}

In the preceding discussion of the Toda chain we saw that the
ratios\footnote{It is perhaps helpful to recall at this point
that theta-functions are used to embed a curve into some projective
space giving the {\em inhomogeneous} co-ordinates of the embedding.
Their ratios
may be considered as ``normal" or {\em homogeneous} co-ordinates.}
 of the theta functions $\Theta_k$ Poisson commuted by appealing to
the explicit form of the equations of motion. In fact a more general result
holds for the Calogero-Moser family that we shall now describe. We have seen
that the theta functions for the Calogero-Moser family satisfy (\ref{thetsim}).
We wish to show that for appropriate solutions (and for all $k,l,m,n$)
\begin{equation}
\label{posscom}
0=\left\{ {\Theta_k\over\Theta_l},
{\Theta_{m}\over\Theta_{n}}\right\}
\Longleftrightarrow
\Theta_l\, S_{kmn}=\Theta_{k}\, S_{lmn}
\end{equation}
where
$$S_{lmn}\equiv\Theta_l\{\Theta_m,\Theta_n\}+\Theta_m\{\Theta_n,\Theta_l\}+
\Theta_n\{\Theta_l,\Theta_m\}=S_{mnl}=-S_{lnm}.$$
Indeed (upon setting $l=n$ here and using the antisymmetry of $S$) we see
that (for all $k,m,n$) $S_{kmn}=0$ or
$$
0=S_{kmn}=\sum_{a=1}\sp{N}
\left| \begin{array}{ccc} \Theta_{k}&\Theta_{m}&\Theta_{n}\\
\frac{\partial\Theta_{k}}{\partial q_a}&\frac{\partial\Theta_{m}}{\partial q_a}
&\frac{\partial\Theta_{n}}{\partial q_a}\\
\frac{\partial\Theta_{k}}{\partial p\sp{a}}&\frac{\partial\Theta_{m}}
{\partial p\sp{a}} &\frac{\partial\Theta_{n}}{\partial p\sp{a}}
\end{array}\right|.
$$
Such addition formulae are very restrictive \cite{Braden} and closely connected
with integrable systems.

We establish for the Calogero-Moser system the commutativity (\ref{posscom})
in the following way \cite{M99c}.
According to \cite{KriCal} the equation
\be
0 = \Theta ({\bf z}|T)
= \sum_{i\in{\Z }_N}\theta_{i\over N}(z| N\tau)\Theta_{i}
\ee
(as an equation on the $z$-torus)
has exactly $N$ zeroes ${z\over N}=q_1,\dots,q_k$. As a
consequence one gets a system of linear equations
\be
\label{systhet}
\sum_{i=1}^N\theta_{i\over N}(Nq_j| N\tau)\Theta_{i} = 0
\ \ \ \ \ \ j=1,\dots,N.
\ee
The system should have nontrivial solutions, i.e.
$\det_{ij}\theta_{i\over N}(Nq_j| N\tau) = 0$, which effectively reduces the
number of degrees of freedom from $N$ to $N-1$.
Then (\ref{systhet}) can be
rewritten as
\be
\sum_{{i=1}\atop{i\neq i_0}}^{N-1}\theta_{i\over N}(Nq_j| N\tau)
{\Theta_{i}\over\Theta_{i_0}} = \theta_{i_0\over N}(Nq_j| N\tau),
\qquad
\forall i_0; \quad j=1,\dots,N-1,
\ee
and so using Cramers rule
\begin{equation}
\label{sollisy}
{\Theta_{i}\over\Theta_{i_0}} = {\det_{k\neq i_0,i\to i_0;j=1,\dots,N-1}
\theta_{k\over N}(Nq_j| N\tau)\over\det_{k\neq i_0;j=1,\dots,N-1}
\theta_{k\over N}(Nq_j| N\tau)}.
\end{equation}
Therefore, the ratios ${\Theta_{i}\over\Theta_{j}}$ depend only on the
co-ordinate $q_k$, $k=1,\dots,N$ of the Calogero-Moser particles and so
obviously Poisson commute with
respect to the Calogero-Moser symplectic structure
\be
\label{sympfor2}
\Omega^{CM} = \sum_{i=1}^{N}d{q}_i\wedge dp_i
= \sum_{i=1}^{N}d{z}_i\wedge da_i
\ee
restricted to $\sum_{j=1}^Nq_j = const$. The latter condition comes from
the of vanishing the determinant
$$
\det_{ij}\theta_{i\over N}(Nq_j| N\tau) = 0.
$$
Indeed, using the $\theta$-function identities described earlier
coming from Wick's theorem
\cite{Wick,Rui}
\be
\label{wick}
\det_{ij}\theta_{i\over N}(Nq_j| N\tau) \sim
\theta_{\Sigma {i\over N}}\left(N\sum_k q_k | N\tau\right)
\prod_{i<j}\theta_{\ast}(Nq_i-Nq_j| N\tau)
\ee
and we can can compute (\ref{sollisy}) explicitly; the vanishing of the
determinant corresponds to $\sum_{j=1}^Nq_j$ being a zero of the
theta function $\theta_{\Sigma {i\over N}}$ (with characteristic being the
sum of the characteristics ${i\over N}$).
This centre of mass constraint is also equivalent to (\ref{suma}),
$\sum_{j=1}^Na_j =const$.
A consequence of the the ratios ${\Theta_{i}\over\Theta_{j}}$ depending only
on the co-ordinates of the integrable system is that they may be used
to construct a set of independent hamiltonians for a dual system
\cite{BMMM3,M99c}.

\subsection{The Perturbative
Limit of the Calogero-Moser Models}

The elliptic Calogero-Moser model degenerates in the perturbative
limit of the SW theory $\tau\to +i\infty$ giving rise to the well-known
trigonometric Calogero-Moser-Sutherland model. The solution in terms of
the action-angle variables is a
direct generalisation of the open Toda chain case and may be presented in
terms of the dual rational Ruijsenaars-Schneider model. The salient
features are as follows.

The Lax operator
\be
{\cal L}_{ij} =p_i\delta_{ij} + m(1-\delta_{ij}){1\over\sinh q_{ij}}
\ee
may be considered as a limiting case of the Lax operator of the trigonometric
Ruijsenaars-Schneider model (\ref{TRL1}) in the same way as we derived
the Lax operator for the elliptic Calogero-Moser model from its 
Ruijsenaars-Schneider counterpart.
The spectral curve for the model is a minor modification of (\ref{optocu}).
Indeed, from (\ref{dhph}), in the limit $\tau\to\infty$ one gets
(for appropriate imaginary period)
\be
{\sinh\2\left(z-m{\d\over\d\lambda'}
\right)\over\sinh(\2 z)}\ P_N(\lambda' ) = 0.
\ee
Upon introducing $w=e^{z}$, one may express this as (\ref{triCa})
\be
\label{tricacu}
w = {P_N(\lambda)\over P_N(\lambda -m)} = \prod_{i=1}^N{\lambda-a_i\over
\lambda-a_i-m}.
\ee
Again for $su(N)$ we have $\sum a_i = 0$. The spectral curve
is equipped with a generating differential $dS = \lambda {dw\over w}$.

The period matrix in this case is given by \cite{DHPhcu}
\be
\label{trcapema}
-i\pi T_{jk}\sp{Pert.} = -i\pi {\tau}\delta_{jk} +
\2\delta_{jk}\sum_{r\ne j}\log {a_{rj}^2\over (a_{rj}+m)(a_{rj}-m)}
-\2(1-\delta_{jk})\log {a_{jk}^2\over (a_{jk}+m)(a_{jk}-m)}
\ee
where the first term corresponds to the bare coupling of the elliptic
Calogero-Moser model, the modulus of the base torus
$\tau = \theta + {i\over g_{YM}^2}$. In the perturbative limit
$\tau\to +i\infty$ (\ref{trcapema}) this is renormalised and remains
finite. In this limit one gets from (\ref{decfin})
\be
\Theta ({\bf z}|T) = \sum_{k=0}^N e^{2\pi ik z\over N}e^{i\pi k^2\tau\over N}
{\bar\Theta}_k
\ee
so that, finally, the ``tau-functions" of the trigonometric Calogero models may
be introduced as
\be
\label{taucal}
\bar \Theta_k = \sum_{|I|=k}\prod_{i\in I}e^{2\pi iz_i}{\prod_{j\in{\bar I}}}
f(a_{ij}), \qquad
f(a) = \sqrt{1 - {m^2\over a^2}}.
\ee

\section{The Strong coupling limit}

\subsection{Solitonic solutions of the periodic Toda chain }

The quantum moduli space of the 4D pure $su(N)$ \N2 SYM has $N$
maximally singular points at which $N-1$ monopoles become simultaneously
massless. These are the confining vacua of an \1N theory
\cite{SW,DouShe,HaZaS}. At these points the dual variables $a_i\sp{D}$ are
the appropriate variables to describe the prepotential.
The \N2 spectral curve (\ref{fsc-Toda}) at these points is described
in terms of a Chebyshev polynomial $P_N^{\rm Chebyshev}(\lambda)$
defined by $P_N^{\rm Chebyshev}(2\cos v)=2\cos(N v)$.
With $w=e^z$ and this choice of polynomial we see (\ref{fsc-Toda}) turns into
\be\label{chebyshev}
P_N^{\rm Chebyshev}(\lambda)= 2\cosh z.
\ee
From the definition of the Chebyshev polynomial it is clear
$\lambda=2\cosh\left({z\over N}\right)$ is a solution of (\ref{chebyshev}) as
indeed is $\lambda=2\cosh\left({z\over N}+i{2\pi k\over N}\right)$ ($k=0,\ldots
N-1$). These are the \1N points of the theory, related by a $\Z_N$
symmetry, and we will focus on the first of these in performing our analysis.

The hyperelliptic form (\ref{defytoda}) of the spectral curve (recall $y=w-1/w$)
is now
\be
\label{solpeto}
y^2 = P_N(\lambda)^2 - 4 = (\lambda^2-4)Q(\lambda )^2
%=4\sinh^2{z\over N}Q(\lambda )^2
\ee
where the roots of polynomial $Q(\lambda )$ are given by
\be
Q(\lambda ) = \prod _{j=1}^{N-1}(\lambda - 2\cos{\pi j\over N}).
\ee
This is a ``solitonic" curve in the periodic Toda chain: if we express the curve
as $y^2 = \prod _{j=1}^{2 N+2}(\lambda - e_k)$ we see that
$e_{2k}=e_{2k+1}=\cos{\pi k/N}$ ($k=1,\ldots N-1$)
and the corresponding $\bf B$ periods have
collapsed; $e_1=2=-e_{2N+2}$ are single branch points.

Let us now introduce  a new variable by
\begin{equation}
\label{mapcyl}
\lambda = 2\cosh {z\over N} \equiv \xi + \xi ^{-1}.
\end{equation}
Now (\ref{mapcyl}) maps the 2-sheeted cover of the $\lambda$-plane
$y=\sqrt{\lambda^2-4}$ to a cylinder with co-ordinate $\xi$.
Thus eqs.~(\ref{chebyshev}), (\ref{solpeto}) describe analytically a cylinder
with $N-1$ distinguished pairs of points.
With these coordinates our differentials on the curve now take the form
\be
\label{holsol}
{\lambda^{k-1}d\lambda\over y} = {\lambda^{k-1}d\lambda\over
\sqrt{\lambda^2-4}\ \prod _{j=1}^{N-1}(\lambda - 2\cos{\pi j\over N})} =
{d\xi \xi^{N-2}\left(\xi + {1\over\xi}\right)^{k-1}\over
\prod _{j=1}^{N-1}\left(\xi - e^{i\pi j\over N}\right)
\left(\xi - e^{-{i\pi j\over N}}\right)}, \qquad k=1,\dots,N-1.
\ee
For a non-degenerate curve these were holomorphic but now
they acquire simple poles at the points $\xi_j^+ = e^{i\pi j\over N}$ and
$\xi_j^- = e^{-{i\pi j\over N}}$ (and have singularities at the
infinities of the cylinder (\ref{mapcyl}) $\xi = 0,\infty$). The
canonical basis in the space of differentials (\ref{holsol}) can be chosen
as ($j=1,\ldots,N-1$)
\be
d\omega^D_j = {\sin{\pi j\over N}\over\pi}{d\xi\over
\left(\xi - e^{i\pi j\over N}\right)\left(\xi - e^{-{i\pi j\over N}}\right)}
\equiv {\sin{\pi j\over N}\over\pi}{d\xi\over
\left(\xi - \xi_j^+\right)\left(\xi - \xi_j^-\right)} = \\ =
{1\over 2\pi i}\left({d\xi\over\xi - \xi_j^+} - {d\xi\over\xi - \xi_j^-}
\right) = {1\over 2\pi i}d\log{\xi - \xi_j^+\over\xi - \xi_j^-}.
\ee
These differentials are normalised to the ${\bf B}$-cycles, here the cycles
around the marked points $\xi_j^\pm$,
\be
\oint_{B_i}d\omega^D_j =  \oint_{\xi_i^+}d\omega^D_j
=- \oint_{\xi_i^-}d\omega^D_j = \delta_{ij},
\ee
while certain of the ${\bf A}$-periods ($\oint_{A_j}d\omega^D_j =
\int_{\xi_j^-}^{\xi_j^+}d\omega^D_j$) diverge, the others ($j\ne k$)
being given by
\be
\label{invt}
T^D_{jk} = \oint_{A_j}d\omega^D_k = {1\over 2\pi i}\log{\sin^2{\pi\over 2N}(j-k)\over
\sin^2{\pi\over 2N}(j+k)} =
{1\over i\pi}\log{\sin{\pi\over 2N}|j-k|\over
\sin{\pi\over 2N}(j+k)}.
\ee
Using this expression one may show that $T^D$ (\ref{invt}) satisfies
the Edelstein-Mas \cite{ema} identity
\be
\label{mas}
\sum_{k=1}^{N-1}\sin {\pi ki'\over N}\sin {\pi kj'\over N}
\sum_{i,j=1}^N \tilde T^{\rm pert}_{ij}(a_l\to 2\cos{\pi (l-\2)\over N})
\cos{\pi k(i-\2)\over N}
\cos{\pi k(j-\2)\over N}\ =\frac{  N^2}{4} T^D_{i'j'}.
\ee
This conjecture of Edelstein and Mas is proven in Appendix B.
We note that an equivalent expression to (\ref{invt}) was also estabilished
in \cite{Marino} where an interesting investigation of the \1N degenerations
of the ``multisoliton" solutions is presented.

The Abel map in the present setting is
\be
\label{abelsol}
z_j^D = \sum_{k=1}^{N-1}\int^{\xi_k} d\omega_j^D = {1\over 2\pi i}
\sum_{k=1}^{N-1}\log{\xi_k - \xi_j^+\over\xi_k - \xi_j^-} \equiv
{1\over\pi}\sum_{n=1}^\infty t_n\sin{\pi jn\over N}
\ee
where ${\Bf \xi}=\{\xi_k\}$, $k=1,\dots,N-1$ is the set of $N-1$ points
which define the Abel map
\footnote{The Abel map here depends on the genus $N-1$ of the smooth curve
(\ref{fsc-Toda}) that arises from the double scaling limit of the
genus $N$ elliptic Calogero-Moser curve. It is interesting to point out that in
the perturbative limit the basis of differentials (\ref{degholdif}) of the
perturbative curve remembers even more --
that the original curve came from the elliptic Calogero-Moser model and
had genus $N$.}.
The asymptotics at ``infinity" are given by
\be
z_j^D\stackreb{\bxi\to\infty}{=}{1\over 2\pi i}\sum_k
\int^{\xi_k} d\log{\xi - \xi_j^+\over\xi -\xi_j^-}=
{1\over 2\pi i}\sum_k\left(\log\left(1-{\xi_j^+\over\xi_k}\right) -
\log\left(1-{\xi_j^-\over\xi_k}\right)\right)
\\ = -{1\over 2\pi i}\sum_{n=1}^\infty\left((\xi_j^+)^n-
(\xi_j^-)^n\right)\sum_k{1\over n\xi_k^n} =-{1\over\pi}
\sum_{n=1}^\infty \sin{\pi jn\over N}\sum_k{1\over n\xi_k^n}
\ee
and
\be
\label{zddef}
z_j^D\stackreb{\bxi\to 0}{=}{1\over 2\pi i}\sum_k\int^{\xi_k}
d\log{\xi - \xi_j^+\over\xi -\xi_j^-}={1\over 2\pi i}\sum_k\int^{\xi_k}
d\log{1 - {\xi\over\xi_j^+}\over 1 - {\xi\over\xi_j^-}} =
{1\over 2\pi i}\sum_k\left(\log\left(1-{\xi_k\over\xi_j^+}\right) -
\log\left(1-{\xi_k\over\xi_j^-}\right)\right)
\\ = {1\over 2\pi i}\sum_{n=1}^\infty\left((\xi_j^+)^n-
(\xi_j^-)^n\right)\sum_k{\xi_k^n\over n} = {1\over\pi}
\sum_{n=1}^\infty \sin{\pi jn\over N}\sum_k {\xi_k^n\over n}
\ee
provided
\be
t_n\stackreb{\bxi\to\infty}{=}-\sum_k{1\over n\xi_k^n},\qquad
t_n\stackreb{\bxi\to 0}{=}\sum_k{\xi_k^n\over n}.
\ee
We observe that the relation (\ref{abelsol}) with $t_{n\ge2}=0$,
\be
z^D_j = t_1\sin{\pi j\over N}
\ee
coincides with the vacuum value of the string tension, or monopole condensate,
proportional to the SUSY breaking parameter $t_1$.

Naively, in the strong coupling limit ${\tilde T}_{ij}\to 0
$ (and $T^D_{ii}\to\infty $) the $\Theta$-functions (\ref{Theta}) turn into
\be
\Theta_k \stackreb{\tilde T_{ij}\to 0}{=}\
\sum_{\sum_{j=1}^N n_j=k} e^{2\pi i\sum n_j\tilde z_j}
\ee
which is not a well-defined object.
The resolution to this may be seen by considering the $N=2$ case.
Here one appears to have
\begin{equation}
\begin{array}{rl}
\Theta_0 &= \sum_{n_1+n_2=0}e^{2\pi i(n_1\tilde z_1+n_2\tilde z_2)} =
\sum_n e^{2\pi in(\tilde z_1-\tilde z_2)} \stackreb{\tilde z_1=-\tilde
z_2\equiv {Z\over 2}}{=}\ \sum_n e^{2\pi inZ} = \theta_3(Z|\hat T\to 0)
\\
\Theta_1 &= \sum_{n_1+n_2=1}e^{2\pi i(n_1\tilde z_1+n_2\tilde z_2)} =
\sum_{n_1+n_2=1} e^{i\pi (n_1-n_2)Z} \stackreb{n_1-n_2=n\ \rm odd}{=}\
\sum_{n\ \rm odd} e^{i\pi nZ} = \theta_2(Z|\hat T\to 0)
\end{array}
\end{equation}
and, naively
\be
e^q ={\Theta_1\over\Theta_0} = {\theta_2(Z|\hat T\to 0)\over
\theta_3(Z|\hat T\to 0)} = 2\cos \pi Z.
\ee
This appears to have nothing in common with the desired solitonic solution
given by integrating the equations of motion
\be
\int dt = \int {dq\over\sqrt{h-e^{2q}-e^{-2q}}} \stackreb{h=2}{=}
i\int {dq\over e^{q}-e^{-q}}
\ee
in the solitonic limit $h\to 2$, which gives rise to
\be
e^q = i \tan(t-t_0)
\ee
(for an appropriate constant of integration $t_0$).
However the correct answer does appear after both a shift $Z=\bar Z-\2$
and a {\em modular transform} $\hat T\to-{1\over\hat T}\equiv T^D$.
Then one has
\be
e^q = {\Theta_1\over\Theta_0} =  {\theta_1(\bar Z|\hat T\to 0)\over
\theta_4(\bar Z|\hat T\to 0)}
 = i
{\theta_1(\bar Z^D|T^D)\over\theta_2(\bar Z^D|T^D)} \stackreb{T^D\to\infty}{=}\
i\tan\pi \bar Z^D + {\cal O}(e^{2\pi iT^D})
\ee
where ${\bar Z}^D={\bar Z\over\hat T} = - T^D\bar Z={ Z\over\hat T}
+{1\over2\hat T}$.

The combination of both a shift and a modular transformation appears
to work in general. Using the modular properties of theta-functions we have
\begin{equation}
\begin{array}{rl}
\Theta_k (\hat{\bf z}|\hat T) &= \sum_{{\bf m}\in\Z^{N-1}}
e^{2\pi i({\bf m}-{k\over N}{\bf e})\hat{\bf z} + i\pi
({\bf m}-{k\over N}{\bf e})\hat T({\bf m}-{k\over N}{\bf e})}
\\
&= e^{-2\pi i{k\over N}{\bf e}\hat{\bf z} + i\pi
{k^2\over N^2}{\bf e}\hat T{\bf e}}\sum_{{\bf m}\in\Z^{N-1}}
e^{2\pi i{\bf m}(\hat{\bf z}-{k\over N}\hat T{\bf e}) + i\pi
{\bf m}\hat T{\bf m}} = e^{-2\pi i{k\over N}{\bf e}\hat{\bf z} + i\pi
{k^2\over N^2}{\bf e}\hat T{\bf e}}
\Theta \left(\left.\hat{\bf z}-{k\over N}\hat T{\bf e}\right|\hat T\right)
\\
&= e^{-2\pi i{k\over N}{\bf e}\hat{\bf z} + i\pi
{k^2\over N^2}{\bf e}\hat T{\bf e}-i\pi(\hat{\bf z}-{k\over N}\hat T{\bf e})
{1\over\hat T}(\hat{\bf z}-{k\over N}\hat T{\bf e})}\
\left({\det\hat T}\right)\sp{-\2}
\Theta \left(\left.{\hat T}^{-1}\hat{\bf z}-{k\over N}{\bf e}\right|
- {\hat T}^{-1}\right) .
\end{array}
\end{equation}
Thus
\be
{\Theta_k\over\Theta_{k-1}} =
{\Theta ({\hat T}^{-1}\hat{\bf z}-{k\over N}{\bf e}|- {\hat T}^{-1})\over
\Theta ({\hat T}^{-1}\hat{\bf z}-{k-1\over N}{\bf e}|- {\hat T}^{-1})}
=
 e^{i\pi \over N}
{\Theta\left[{{\bf e}\atop0}\right] \left(
{\hat T}^{-1}({\bf z}+\2{\bf e})-{k\over N}{\bf e}|- {\hat T}^{-1}\right)\over
\Theta\left[{{\bf e}\atop0}\right]\left(
{\hat T}^{-1}({\bf z}+\2{\bf e})-{k-1\over N}{\bf e}|- {\hat
T}^{-1}\right)}\equiv \\
\equiv
e^{i\pi \over N}
{\Theta\left[{{\bf e}\atop0}\right] \left(
\hat{\bf z}^D-{k\over N}{\bf e}|- {\hat T}^{-1}\right)\over
\Theta\left[{{\bf e}\atop0}\right] \left(
\hat{\bf z}^D-{k-1\over N}{\bf e}|- {\hat T}^{-1}\right)}
\ee
where now
$$
\hat{\bf z}^D={\hat T}^{-1}({\bf z}+\2{\bf e})
$$
and ${\bf e}\equiv(1,\dots,1)$.
The effect of the shift is to keep only the leading terms in the quadratic
$({\bf m}+\2{\bf e}){\hat T}^{-1}({\bf m}+\2{\bf e})$ in the limit
$\hat T\to 0$ yielding
\be
\label{thetsol}
{\Theta_k\over\Theta_{k-1}}
\stackreb{\hat T\to 0}{=}  e^{i\pi \over N}
{\vartheta_k(\hat{\bf z}^D)\over\vartheta_{k-1}(\hat{\bf z}^D)}
\ee
where
\be
\vartheta_k(\hat{\bf z}^D) = \sum_{{\bf s}\in\Z_2^{N-1}}e^{i\pi {\bf s}
({\bf z}^D-
{k\over N}{\bf e})-{i\pi\over4}\sum_{j\neq j'}s_j{\hat T}^{-1}_{jj'}s_{j'} }
\ee
and $\Z_2\equiv\{+1,-1\}$.

Using (\ref{zddef}), (\ref{thetsol}) and identifying the off diagonal
parts of ${\hat T}^{-1}$ with $T^D$ one can propose formulae for the
Baker-Akhiezer functions
\be
\Psi_k (\xi,t) \sim \xi^k e^{\sum_n t_n(\xi^n-\xi^{-n})}
{\vartheta_k(\hat{\bf z}^D(t_n;\xi))\over\vartheta_k(\hat{\bf z}^D(t_n))}
\ee
where
\be
\label{tausol}
\vartheta_k(\hat{\bf z}^D(t_n)) = \sum_{m_j=\pm}e^{-i\pi{k\over N}\sum_jm_j+
i\sum_{j,n} m_jt_n\sin{\pi jn\over N}+{1\over4}\sum_{j\neq j'}m_jm_{j'}
\log{\sin{\pi\over 2N}(j+j')\over\sin{\pi\over 2N}|j-j'|}}
\ee
and
\be
\vartheta_k(\hat{\bf z}^D(t_n;\xi)) = \sum_{m_j=\pm}e^{-i\pi{k\over N}\sum_jm_j+
i\sum_{j,n} m_jt_n\sin{\pi jn\over N}+\sum_{j\neq j'}m_jm_{j'}
\log{\sin{\pi\over 2N}(j+j')\over\sin{\pi\over 2N}|j-j'|}}
\prod_j\left({\xi - \xi_j^-\over\xi -\xi_j^+}\right)^{m_j/2}
\\
= \sum_{m_j=\pm}e^{-i\pi{k\over N}\sum_jm_j+
i\sum_{j,n} m_jt_n\sin{\pi jn\over N}}\prod_j\left({\xi - e^{-{i\pi j\over N}}\over\xi -
e^{i\pi j\over N}}\right)^{m_j/2}\prod_{j\neq j'}
\left({\sin{\pi\over 2N}(j+j')\over\sin{\pi\over
2N}|j-j'|}\right)^{m_jm_{j'}\over4}
.
\ee
Low order cases of this are
\begin{itemize}
\item $SL(2)$. $N=2$, $N-1=1$, $m_j=m_1=m=\pm$. Then
\be
\vartheta_k(\hat{\bf z}^D(t_n;\xi))= \sum_{m=\pm}e^{-{i\pi km\over 2} +imt}
\left({\xi + i\over\xi - i}\right)^{m/2} = {1\over\sqrt{\xi^2+1}}
\left( e^{i(t-{\pi k\over 2})}(\xi+i)+e^{-i(t-{\pi k\over 2})}(\xi-i)\right)
\ee
\item $SL(3)$. $N=3$, $N-1=2$, $j=1,2$, $(m_1,m_2)=\{ (++),(+-),(-+),(--)\}$.
\be
\label{ztsl3}
\vartheta_k(\hat{\bf z}^D(t_n;\xi)) = \sum_{m_1,m_2=\pm}
e^{-{i\pi k\over 3}(m_1+m_2) +i(m_1z_1+m_2z_2)}
\left({\xi - e^{-{i\pi\over 3}}\over\xi - e^{i\pi\over 3}}\right)^{m_1/2}
\left({\xi - e^{-{2\pi i\over 3}}\over\xi - e^{2\pi i\over 3}}\right)^{m_2/2}
2^{m_1m_2\over2}
\\
z_1=\sum t_n\sin{\pi n\over 3},\qquad
z_2=\sum t_n\sin{2\pi n\over 3}.
\ee
\end{itemize}
These formulae give rise to the general form for the solitonic
Baker-Akhiezer function of the periodic Toda chain
\be
\Psi_k (\xi,t) = \xi^ke^{\sum_j t_j(\xi^j - \xi^{-j})}
{R_k (\xi,t)\over R(\xi)}.
\label{basol}
\ee
Here $R(\xi )$ is a normalisation factor, independent of times, and chosen
to be a polynomial of $\xi $ of degree $N-1$ in order for (\ref{basol}) to
have desired analytic properties, while
\be
R_k(\xi,t)=\psi_k(t)\prod _{s=1}^{N-1}(\xi - \mu_s(k,t)) =
\sum_{l=0}^{N-1}r_l(k,t)\xi^l .
\ee
These Baker-Akhiezer functions are defined for $k:0\ldots N-1$ and extended to
all $k$ by
\be
\Psi_{k+N} = w\Psi_k,\qquad
w = \xi^N
\label{bascdef}
\ee
(equivalently $R_{k+N} = R_k$).
Now the Toda chain Lax equation
\be
\lambda\Psi_n = C_{n+1}\Psi_{n+1} + p_n\Psi_n + C_n\Psi_{n-1},\qquad
C_n \equiv e^{\2(q_n-q_{n-1})}, \qquad
\lambda = \xi + {1\over\xi}
\ee
implies that
\be
r_0(n) - C_nr_0(n-1) = 0
\\
r_1(n) - C_nr_1(n-1) - p_nr_0(n) = 0
\ee
and so
\be
r_0(n) = C_nr_0(n-1) = \dots = e^{\2(q_n-q_0)}r_0(0) \sim e^{\2q_n}.
\ee
For the solitons coming from degeneration of $N$-periodic Toda chain
one should also impose the ``gluing conditions"
\be
\Psi_n (\xi_j) = \Psi_n({1\over\xi_j})\ \ \ \ \ \ \ j=1,\dots,N-1.
\label{glue}
\ee
This means that the Baker-Akhiezer function remembers that it
originally came from a
genus $N-1$ Riemann surface, and each pair of points $\xi_j, {1\over\xi_j}$
corresponds to a degenerate handle. Now the conditions (\ref{glue}) together
with (\ref{bascdef}) entail
\be
\xi_j^{2N} = 1.
\ee
Thus we may take
\be
\xi_j = e^{i\pi j\over N}
\ee
where the label $j$ can be restricted to $j=1,\dots,N-1$ since
\be
\phi_j = \xi_j + {1\over\xi_j} = 2\cos{\pi j\over N} = \phi_{2N-j}.
\label{vacua}
\ee
Eq.~(\ref{glue}) explicitly reads
\be
{R_n({1\over\xi_j})\over R_n(\xi_j)} =
\prod_{k=1}^{N-1}{\xi_j^{-1}-\mu_k(n,t)\over\xi_j - \mu_k(n,t)} =
e^{{2\pi inj\over N}+4i\sum_l t_l \sin{\pi jl\over N}+ Z_j(R)},
\qquad
j=1,\dots,N-1.
\label{glue2}
\ee
Here $Z_j(R)=\log{R({1\over\xi_j})\over R(\xi_j)}$, thus if on
has chosen $R(\xi )=\prod_{s=1}^{N-1}(\xi-\gamma_s)$ in (\ref{basol}) then
$Z_j(R)=\sum_{s=1}^{N-1}\log{{{1\over\xi_j}-\gamma_s}\over {\xi_j-\gamma_s}}$.
Now (\ref{glue2}) is a system of linear equations for the coefficients
$r_k(n,t)$ of the polynomial $R_n(\xi,t)$:
\be
\sum_{k=0}^{N-1}
\sin\left({\pi jk\over N}+{\pi jn\over N}+2\sum_l t_l \sin{\pi jl\over N}
-{i\over2}Z_j(R)\right) r_k(n,t)=0.
\ee
Such can be readily solved.
Conditions (\ref{vacua}) can be interpreted as values of the
scalar fields at the critical points of the superpotential, while the soliton
trajectories connect the critical points.

\subsection{On the ``solitonic" limit of the elliptic Calogero-Moser system}

The solitonic limit of the elliptic Calogero-Moser system involves many open
question, though it seems that we can proceed in an analogous manner to the
Toda chain case. We will discuss here only some explicit computations for low N
and will postpone the computation of the soliton phases or string tensions.

For the $N=2$ case the solitonic limit corresponds to
$h = e = \wp (\omega)$ ($\omega$ may
be any (!) half-period) so that (\ref{cdhph2}) turns into
\be
\label{solcan2}
\lambda^2 - x + h = \lambda^2 - \wp (z) + \wp (\omega) =
\left(\lambda - \Phi(z,\omega)\right)\left(\lambda + \Phi(z,\omega)\right) = 0.
\ee
Thus
\be
\lambda = \pm \Phi(z,\omega) = \pm
{\sigma(z-\omega)\over\sigma(z)\sigma(\omega)}\ e^{\zeta(\omega)z} =
\mp {\sigma(z+\omega)\over\sigma(z)\sigma(\omega)}\ e^{-\zeta(\omega)z}
\ee
and these functions are related to the entries of the Calogero-Moser Lax
operator (\ref{LaxCalma}) at half-periods via
\be
F(\omega|z) = {\sigma(z+\omega)\over\sigma(z)\sigma(\omega)} =
\Phi(z,\omega)e^{\zeta(\omega)z}
\\
F(-\omega|z) = - {\sigma(z-\omega)\over\sigma(z)\sigma(\omega)}
={\sigma(z+\omega)\over\sigma(z)\sigma(\omega)}e^{-2\zeta(\omega)z} =
\Phi(z,\omega)e^{-\zeta(\omega)z}.
\ee
In this limit one also has
\be
y^2 = (x-e)(x-e_+)(x-e_-) = \lambda^2(\lambda^2-\tilde e_+)
(\lambda^2-\tilde e_-),
\qquad
\tilde e_\pm = e_\pm - e.
\ee
The Seiberg-Witten differential and periods are
\be
dS = \lambda{dx\over y} = {dx\over \sqrt{(x-e_+)(x-e_-)}},
\qquad
\oint_B dS = 0,
\qquad
\oint_A dS = 2\pi i .
\ee
Thus instead of $\Sigma^{CM}_{\rm reduced}$ (\ref{calN2red}) one may
introduce the {\em rational} reduced curve
\be
\label{rason2}
Y^2 = (x-e_+)(x-e_-).
\ee
Indeed, a direct integration of the equation of motion (\ref{intred}) gives
now
\be
2 i t = \int {dx\over (x-e)\sqrt{(x-e_+)(x-e_-)}},
\ee
a contour integral on the {\em rational} curve (\ref{rason2}).

Let us note that formula (\ref{solcan2}) can be obtained by considering the
extrema of the Calogero-Moser hamiltonian (the superpotential in the SW
approach). Then
\be
dh = {\d h\over\d p}dp + {\d h\over\d q}dq = 0
\ee
yields
\be
p=0,
\qquad
\wp'(q)=0.
\ee
The latter is satisfied by {\em any} half-period $q=\omega$, so that
\be
\left. h\right|_{dh=0} = \wp(\omega) = e.
\ee
A similar argument holds for the general $N$-particle case. Now
the Lax operator (\ref{LaxCal}) computed at the ``stationary" points yields
$p_1=\dots=p_N=0$ and the locus equation
\be
\label{dwp}
{\cal L}_N=\{ q_j\big|\, q_j\ne q_i,\
 \sum_{j\neq i}\wp'(q_{ij}) = 0, \ i:1\ldots N\}.
\ee
The (closure of) this locus has a rich geometry and many questions
regarding it are still unanswered. The early work of \cite{AMM} is still
one of the most detailed investigations (see also \cite{vaninsky}).
The (closure of) the locus has in general several disconnected pieces some
of which are trivial copies of the torus. The latter are easily understood:
any odd, periodic function (with period $L$) satisfies the identity
\be
\label{perod}
f\left({L\over N}\right) + f\left({2L\over N}\right) +
f\left({3L\over N}\right) + \dots + f\left({N-2\over N}L\right)
+ f\left({N-1\over N}L\right) = 0
\ee
for any odd $N$ and  for $N$ even provided
$f\left({L\over 2}\right)=0$.
Choosing $\wp'(x)$ as the function $f(x)$ with
$L=2\omega$ (and $\wp'(\omega)=0$) one gets a solution to (\ref{dwp}) with
\be
q_k = q_0 + {2\omega\over N}k,
\qquad
q_{jk} = {2\omega\over N}(j-k).
\ee
The locii here are simple copies of the torus; by varying the periods in this
construction one gets further simple solutions. There are however other
solutions less well understood.

Lets consider the $N=3$ case.
For $N=3$ eq.~(\ref{dwp}) is equivalent to
\be
\wp'(q_{12}) = \wp'(q_{23}) = - \wp'(q_{13}).
\ee
This has solutions (together with $q_{12}+q_{23}=q_{13}$)
\be
q_{12}=\omega,\ \ \ \ \ q_{23}=\omega',\ \ \ \ \ q_{13}=\omega+\omega'
\ee
and
\be
\label{solso}
q_{12}={2\omega\over 3},\ \ \ \ \ q_{23}={2\omega\over 3},\ \ \ \ \
q_{13}={4\omega\over 3} = - {2\omega\over 3}
\ee
for {\em any} half-period $\omega=(\omega,\omega',\omega+\omega')$
\footnote{One also has the relation between $\zeta$-functions
$\eta=\zeta(\omega)$ in the half-periods
\be
\zeta(\omega+\omega') = \zeta(\omega) + \zeta(\omega')
\ee
}.
Upon substituting (\ref{solso}) and $p_i=0$ into the Lax
equation (\ref{LaxCal}) one gets
\be \lambda^3 + 3F_+F_-\lambda - F_+^3 + F_-^3 = 0 \ee
with the three solutions
\be \label{sol3} \lambda_0 = F_+ - F_-,
\qquad \lambda_\pm = e^{\pm{2\pi i\over 3}}F_+ + e^{\pm{i\pi\over 3}}F_-.
\ee
Here
\be
F_\pm = {\sigma\left(z\pm{2\omega\over 3}\right)\over
\sigma(z)\sigma\left({2\omega\over 3}\right)}e^{\mp{2\over 3}\zeta(\omega)z}.
\ee

\section{Discussion}

In this paper we have considered the singular limits of various
SW integrable systems that are relevant for the weak and strong coupling limits
of the corresponding field theories. We conclude by making several
comments about the relation of these relatively straightforward calculations
in integrable models with corresponding properties of the SUSY gauge theories.

Let us begin with the outstanding problem of why there is a correspondence
at all between integrable systems and SW theory. Part of the
problem is that, from a purely four-dimensional perspective,
Seiberg-Witten theory only sees the commuting Hamiltonians of a mechanical
system. Only these quantities appear as coefficients of the Seiberg-Witten
curve. Half of phase space is not apparent at all and the choice of mechanical
system appears arbitrary. We believe viewing these
systems from a three dimensional perspective sheds light on the matter. In fact
one should also pay attention to the four dimensional \N2 gauge
theories compactified on $\mathbb{R}\sp3\times S\sp1$ \cite{SW3}.
Recall that
the phase spaces of integrable systems play the role of moduli spaces,
or spaces of vacua parameters, of (compactified) SUSY gauge theories in the
following way. The minima of the scalar potential in the gauge theories with
extended supersymmetry
\be
V(\Phi ) = \Tr\sum_{I<J} [\Phi^I,\Phi^J]^2
\label{pot}
\ee
correspond to simultaneously diagonalisable matrices $[\Phi^I,\Phi^J]=0$ whose
eigenvalues $\{ \phi^I_k\}$ can be thought of as the (complexified)
momenta of some ``particles".  Now for theories with a compact dimension
one should also add ``co-ordinates" $\{ q_k\sp{\mathbb R}\}$
corresponding to the eigenvalues of the Wilson loops $\oint A_\mu dx^\mu$.
For \N2 vector multiplets in four dimensions
\footnote{An \N2 4D vector supermultiplet in
the adjoint representation consists of an \1N 4D
vector multiplet $(A_\mu,\psi)$ together with an \1N 4D scalar multiplet
$(\phi, \chi)$.
Here $\psi$ and $\chi$ are two complex Weyl spinors. If say $\chi$
acquires a nontrivial phase (\ref{phase}) under a shift along the loop in the
compact direction,
it becomes massive with the mass ${\epsilon\over R}$. The 4D \1N vector
multiplet remains massless, and can be represented by a 3D \N2
supermultiplet $(A_\alpha,\psi,{q\over R})$, where $\alpha=0,1,2$, $q=RA_3$
and $\psi $ is 3D complex spinor.}
one has a single complex scalar
and compactification of one space-time dimension gives rise to an extra
complex scalar $q=q\sp{\mathbb R}+i\gamma$, where $q\sp{\mathbb R}$
comes the set of eigenvalues of the Wilson
loop  $\oint A_3$ and $\gamma$ corresponds to the (set of) 3D dual photons
$\d_\alpha A_\beta = \epsilon_{\alpha\beta\rho}\d_\rho\gamma$, $\alpha,\beta=
0,\dots,2$.
Thus by viewing the theory on $\mathbb{R}\sp3\times S\sp1$ one may
naturally include coordinates. The moduli space of vacua for the theories
on $\mathbb{R}\sp3\times S\sp1$ is a hyper-Ka\"hler manifold, and, we recall,
such are the phase spaces of algebraically completely integrable systems.
Further, it was argued \cite{SW3} that there is a distinguished
complex structure on this moduli space independent of the radius of the
compact $S\sp1$. This yields the complex structure of the mechanical system.
In general we expect the symplectic structure to arise from a careful
treatment of the central charges of SUSY algebra:
\be
\Omega \sim Z^{BPS} \sim \int_{d\sigma_{jk}}\Tr (F_{jk} + i\tilde F_{jk})\bPhi
\sim
\Tr\int\delta{A}\wedge\delta\Phi \sim \sum \delta q_i\wedge \delta p_i.
\label{bss}
\ee
For a \4N theory, we have {\em three} different choices for the
symplectic structure (\ref{bss}), corresponding to the three different
scalar fields in the \4N theory, and these are related by a
``hyper-K\"ahler" rotation.

Now  consider the decompactification of the ``3D" gauge theory
on $\mathbb{R}\sp3\times S\sp1$ with
symplectic form (\ref{bss}) in the limit of the $S\sp1$ radius $R\to\infty$.
The dimension of the moduli space of the $\mathbb{R}\sp3\times S\sp1$ theory 
is twice that 
of the limiting 4D theory: the ``coordinates" are no longer
present. The resulting 4D gauge theory is associated to
an integrable system: the 4D moduli are, from the 3D point
of view, the Poisson-commuting {\em hamiltonians} constructed from the
3D momenta and co-ordinates, with respect to symplectic structure (\ref{bss}).
From this perspective, quantum effects in the compactified \N2 gauge theory
turn the bare symplectic form (\ref{bss}) into
\be
\sum \delta q_i\wedge\delta p_i \mapsto \sum\delta z_i\wedge\delta
a_i,
\label{qss}
\ee
where the SW integrals \cite{SW}
\be
a_i = \oint_{A_i}dS
\ee
are the correct quantum variables.
One may consider $a_i = a_i(\Phi,\Lambda)$
as a transformation from the bare quantities $\{\phi_i\} $ to
their exact quantum values $\{a_i\}$ playing the role of the  quantum BPS
masses of the effective theory.
In the same way one should consider the
transformation $q_i\rightarrow z_i$ as transformation from bare
values of the monodromy to the exact quantum values of the effective theory.

For theories with four-dimensional \4N SUSY the effective couplings and BPS
masses (i.e. the eigenvalues of the scalar fields $\{\phi_i\}$ and
Wilson loops)
are {\em not renormalised}, since the symmetries of the theory include
the conformal group and so there are no dimensionful scale or mass parameters.
The corresponding integrable model is a system of {\em free} particles:
the Hamiltonians are $u_k = {1\over k}\Tr\Phi^k = {1\over k}\sum_i p_i^k$,
where $\{ p_i\}$ play the role of momenta and the co-ordinates coming from the
``compact" moduli depend linearly on the  angles.
Breaking four dimensional supersymmetry down to \N2 (for example,
 by adding extra mass terms
$m_i^2\Tr\Phi_i^2$ for two of the three scalar fields) reduces the dimension of
the ``scalar" moduli space down to $2N$, that of one complex ``diagonal
matrix" field. Moreover, in contrast to the \4N theory, the
matrices of the scalar fields and the monodromies become dependent upon
each other, or satisfy a nontrivial commutation relation (coming from the
vanishing of D-terms)
\be
[A,\Phi] \sim mJ
\label{momap}
\ee
in the general case (of nontrivial boundary conditions).
Here $J$ is some matrix of ``gauge-covariant" form and the right hand side here
is linear in the parameter of the ``massive deformation" \cite{don, SWCal},
(for $m\to 0$ one comes back to \4N theory).

Let us consider the compactification of an \N2 SUSY Yang-Mills theory with just
a vector supermultiplet to $\mathbb{R}\sp3\times S\sp1$ (with $S\sp1$ having
radius $R$) in more detail.\footnote{
In practice this means for the theory at mass scale
$\Lambda=\Lambda_{QCD}$ that $\Lambda >> {1\over R}$ corresponds to a 4D theory
while $\Lambda << {1\over R}$  to a 3D theory. For the couplings one has
${1\over g_3^2} = {R\over g_4^2}$, i.e. the 3D theory with fixed coupling $g_3$
corresponds to $R\to 0$, or $g_4\to 0$ and so to the weak coupling limit of the
4D theory. In this limit the 4D instantons decouple since the 4D coupling
$g_4$ is small (for $R\to 0$ and $g_3={\rm fixed}$) and so that the
exponential terms, powers of
$e^{-{1\over g_4^2}}$, may be neglected.  In the 3D theory the integral for
the polarization operator
\be
\left(k_\mu k_\nu - \delta_{\mu\nu}k^2\right)\left|{1\over m}\right| + \dots
\ee
yields a convergent integral by dimensional arguments.
In the ``intermediate" case of a compactified (3+1)-dimensional theory
on $\mathbb{R}\sp3\times S\sp1$ with finite $S\sp1$ radius $R$ one
can present this result in terms of a 3D theory together with a sum of
Kaluza-Klein contributions
\be
\label{3+1d}
{1\over g_3^2} = \sum_n \left|{1\over m+{n\over R}}\right|
\ee
For $R\to 0$ only the term with $n=0$ survives in the sum (\ref{3+1d})
leading back to the 3D expression. For the opposite limit $R\to\infty$ one
can define the dimensionless 4D coupling and replace the (divergent)
sum by an integral
\be
{1\over g_4^2} = {1\over Rg_3^2} = {1\over R}\sum_{|n|\le|n_{\rm max}|}
\left|{1\over m+{n\over R}}\right| =
\int_m^{{n_{\rm max}\over R}\equiv\Lambda}{d\mu\over\mu} = \log{\Lambda\over m}.
\ee}
If one takes all the fields to have {\em periodic} boundary conditions in
the compact direction this would yield an \4N (in the the three-dimensional
sense) SUSY theory.  If, however, one puts \cite{BMMM1,BMMM2}
\be
\phi (x+R) = e^{i\epsilon}\phi (x)
\label{phase}
\ee
on {\em half} of the fields, the resulting theory would have only \N2
{\em three}-dimensional SUSY (\1N in the four-dimensional sense),
i.e. the supersymmetry
will be (partially) broken by the non-periodic boundary conditions.
Now in contrast to \4N SUSY in 3D, an \N2 supersymmetric theory can generate a
superpotential \cite{AHW}. In terms of the complexified variables
$q=q\sp{\mathbb R}+i\gamma$, the
superpotential acquires the form \cite{SW3,n23d}
\be
\label{supot34d}
W \sim  \epsilon\Tr\Phi^2 +
{\epsilon\over R^2}\left(\sum_{i=1}^{N-1} e^{q_{i+1}-q_i} +
e^{q_1-q_N}\right).
\ee
Here the first term is the ``4D contribution" while the second term (the first
term in the brackets) has a 3D origin; the final term is induced by 3+1D
instanton contributions. The simple roots of the second term are the usual
3D instantons (4D BPS ``mo\-no\-po\-les") and give the potential of the
{\em open} Toda chain: $\Pi_2\big({SU(N)\over {U(1)\sp{N-1}}}\big)\cong
{\mathbb Z}\sp{N-1}$. The final term (minus the highest root) appears only in
3+1 dimensions and can be treated as a 4D instanton (or caloron) contribution.
In the perturbative limit, considered in detail in this paper, one may
directly see the nonrenormalisability of the superpotential, which means,
in particular, that
\be
W^{4D} = h = a^2 = p^2 + e^{2q} = W^{3D}.
\ee

Now let us discuss how the formulas obtained in this paper may be
be reinterpreted from this three-dimensional perspective. First, we have
discussed the singular limits of the SW integrable systems, when the spectral
curves
degenerate and become rational with marked points (i.e. their ``smooth" genus
is zero -- eqs.~(\ref{polyn}), (\ref{polsu2}), (\ref{triCa}), (\ref{triRu}),
(\ref{tricacu}), (\ref{solpeto}) and (\ref{rason2})). Degeneration means
that the discriminant of the corresponding smooth curve vanishes as one
approaches particular (boundary) points of the moduli space. In the paper
above we have divided these singular limits into two groups: weak-coupling
from the point of view of the SUSY gauge theory which yields ``open" (or
``trigonometric") integrable systems; and strong coupling yielding
``solitonic" behaviour. If one deals with the gauge theory with
{\em extended} supersymmetry each of these degenerate curves as well as
the smooth ones play equivalent roles as physical vacua. By
breaking the extended supersymmetry down to \1N (in the four-dimensional sense)
however, one generates
integrable dynamics {\em distinguishing} the second, ``solitonic", group of
degenerate curves from those of the ``perturbative" ones, so that only the
solitonic points play the role of vacua in \1N theories. It is clear
from the explicit form of solitonic solutions obtained here that the
angles, corresponding to Wilson loops, are proportional to the parameters
of the partial SUSY violation (cf. eqs.~(\ref{abelsol}) and (\ref{ztsl3}) with
refs. \cite{DouShe,HaZaS}).
For both kinds of singular limits we have shown how explicit
solutions may be straightforwardly calculated. The co-ordinates and momenta of
the integrable systems of particles (the classical 3D moduli) are expressed
through the theta-functions of the SW curves. In the singular limits these
theta-functions degenerate into the {\em finite} sums -- see
eqs.~(\ref{thetaop}), (\ref{bartheta}), (\ref{mamo}), (\ref{mamodet}),
(\ref{taucal}), (\ref{tausol}) -- suggestive of a kind of partition function
for a ``discretized" matrix models. The ``dual" point of view is to
consider the theta-functions $\Theta_k$ as generating functions for
Hamiltonians of a dual integrable systems with co-ordinate's $a_i$
and momenta $z_i$. The Poisson commutativity of their ratios, implicitly
proven in \cite{M99c}, can be checked by straightforward calculation in the
degenerate limits.
The fact that degenerate theta-functions appear in the form of ``discretized"
matrix models (\ref{mamo}), where instead of an integral over the real line
one has a (multiple) discrete sum over the {\em spectrum} of the Toda
molecule, may be considered as a possible (though very speculative) sign
of a dual description of SW theory in the language of quantum gravity,
possibly in a ``holographic" sense.

\section*{Acknowledgements}

We are indebted to V.Fock, A.D. Gilbert, S.Kharchev, I.Krichever, 
A.Vainshtein, A.Yung and A.Zabrodin for illuminating discussions
and helpful remarks. A.M. is also grateful
to T.Tomaras and other members of the Theoretical Physics group at the
University of Crete where the essential part of this work has been done.
The work of A.M. was partially supported by RFBR grant No.
00-02-16477, INTAS grant No. 99-01782 and CRDF grant No. RF-2102 (6531) and
grant for the support of scientific schools No. 00-15-96566.
HWB acknowledges a travel grant from the LMS which helped this
collaborations.

\appendix
\section{Elliptic Functions and the Inozemtsev Limit}
In this appendix we shall review some basic definitions and properties
of elliptic functions and then consider the Inozemtsev limit.

The Weierstrass elliptic function $\wp (z|\omega,\omega')$ is the doubly
periodic function with periods $2\omega$, $2\omega'$ given by
\begin{equation}
\wp (z|\omega,\omega') = \frac{1}{z^2}+{\sum_{n,m\in\Z}}'\left(
{1\over (z+2\omega n+2\omega' m)^2} -{1\over (2\omega n+2\omega' m)^2}\right).
\end{equation}
It satisfies the differential equation
\begin{equation}
\wp (z)'\sp2 =4(\wp (z)-e_1)(\wp (z)-e_2)(\wp (z)-e_3),
\label{diffpz}
\end{equation}
and
scaling relation
\begin{equation}
\wp (tz|t\omega,t\omega') =t\sp{-2}\wp (z|\omega,\omega').
\end{equation}
Using
$$
\sin\pi x = \pi x\prod_{n=1}^\infty\left(1-{x^2\over n^2}\right), \qquad
\sum_{n\in\Z}{1\over (x+n)^2} = {\pi^2\over\sin^2\pi x}
$$
we may write
\be
\wp (\tilde z|\2,\frac{\tau}{2})=
\sum_{m\in\Z}{\pi^2\over\sin^2 \pi(\tilde z+m\tau )} - {\pi^2\over 3} -
{\sum _{m\in\Z}}'{\pi^2\over\sin^2\pi m\tau}
\ee
where $\tilde z=\frac{z}{2\omega}$.
A slight rewriting of this utilising the scaling formula yields
\be
\wp (v|i\pi\tau,i\pi)=
\frac{1}{12}-\2\sum_{m>0}\frac{1}{\sinh\sp2(i\pi m \tau)}+
\2\sum_{k=-\infty}\sp{\infty}\frac{1}{\cosh(v+2i\pi k\tau)-1}.
\ee
This is uniformly convergent for $Im(\tau)>0$.

The roots of $e_i$ of equation (\ref{diffpz}) may be expressed in terms of
theta functions by
\begin{equation}
\begin{array}{rl}
e_1 &= {\pi^2\over 12\omega^2}\left(\theta_3(0)^4+\theta_4(0)^4\right) \\
&=
{\pi^2\over 12\omega^2}\prod_{n=1}^\infty(1-q^{2n})^4\left(\prod_{n=1}^\infty(1+
q^{2n-1})^8+ \prod_{n=1}^\infty(1-q^{2n-1})^8\right)
\\
e_2 &= {\pi^2\over 12\omega^2}\left(\theta_2(0)^4-\theta_4(0)^4\right) \\
&= {\pi^2\over 12\omega^2}\prod_{n=1}^\infty(1-q^{2n})^4
\left(16q\prod_{n=1}^\infty (1+q^{2n})^8
- \prod_{n=1}^\infty(1-q^{2n-1})^8\right)
\\
e_3 &= - {\pi^2\over 12\omega^2}\left(\theta_2(0)^4+\theta_3(0)^4\right)\\ &=
- {\pi^2\over 12\omega^2}\prod_{n=1}^\infty(1-q^{2n})^4
\left(16q\prod_{n=1}^\infty(1+q^{2n})^8
+ \prod_{n=1}^\infty(1+q^{2n-1})^8\right)
\end{array}
\end{equation}
where $q=e^{i\pi\tau}$ and $\tau = {\omega'\over\omega}$.
As $q\to0$ we have
$$
e_1 ={\pi^2\over 6\omega^2} + O(q^2),
\quad
e_2= - {\pi^2\over 12\omega^2} + {\pi^2\over 12\omega^2}16q + O(q^2),
\quad
e_3=- {\pi^2\over 12\omega^2} - {\pi^2\over 12\omega^2}16q + O(q^2).
$$
Thus in this limit
\be
e_1\equiv e = {\pi^2\over 6\omega^2}
\\
e_2= e_3 = e_\pm = - {\pi^2\over 12\omega^2}
\\
x = \wp (z) = - {\pi^2\over 12\omega^2} + {\pi^2\over 4\omega^2}{1\over\sin^2\pi
\tilde z}
\equiv {\pi^2\over 4\omega^2}({\tilde x} - {1\over 3})
\\
y = - \2\wp'(z) =  {\pi^3\over 8\omega^3}{\cos\pi\tilde z\over\sin^3\pi\tilde z}
\equiv {\pi^3\over 8\omega^3}{\tilde y}
\\
{\tilde y}^2 = {\tilde x}^2({\tilde x}-1)
\ee

Consider now $m\sp2(\wp(v)-\wp(z))$. The Inozemtsev limit is a double scaling
limit in that the coupling constant $m$ here is scaled as well as the period
of the $\wp$ function. Let
\be
m=M e\sp{-i\pi\tau/2},\qquad v=\phi-i\pi\tau,\qquad w=e\sp{z+i\pi\tau},
\ee
then
\begin{equation}
\begin{array}{rl}
m\sp2(\wp(v)-\wp(z))&=\frac{M\sp2}{2}
\sum_{k=-\infty}\sp{\infty}\left(\frac{e\sp{-i\pi\tau}}{\cosh(\phi+i\pi
 (2k-1)\tau)-1}-
\frac{e\sp{-i\pi\tau}}{\cosh(z+2i\pi k\tau)-1}\right)\\
\\
&{\longrightarrow \atop {Im(\tau)\rightarrow\infty} }2 M\sp2\cosh\phi
-M\sp2\left(w+\frac{1}{w}\right).
\end{array}
\end{equation}

\section{Proof of the Edelstein-Mas Conjecture}

We have
\be
-2i\pi{\tilde T}_{ij}^{\rm pert} =
\delta_{ij}\sum_{l\neq i}\log(a_{i}-a_l)\sp2 -
(1-\delta_{ij})\log(a_{i}-a_j)\sp2 .
\ee
Let us denote by
\be
\label{Sconj}
S=
\sum_{k=0}^{N-1}\sin {\pi ki'\over N}\sin {\pi kj'\over N}
\sum_{i,j=1}^N \tilde T^{\rm pert}_{ij}(a_l\to 2\cos{\pi (l-\2)\over N})
\cos{\pi k(i-\2)\over N}
\cos{\pi k(j-\2)\over N}
\ee
The Edelstein-Mas conjecture is (for $i'\ne j'$)
\be
\label{mass}
S \stackrel{???}{=}\  \frac{N^2}{4} T^D_{i'j'}
\ee
and we shall now establish this.
We have shown that a direct calculation yields (for $j\ne k$)
\be
T^D_{jk} = \oint_{A_j}d\omega^D_k = {1\over 2\pi i}\log{\sin^2{\pi\over 2N}(j-k)
\over
\sin^2{\pi\over 2N}(j+k)} =
{1\over i\pi}\log{\sin{\pi\over 2N}|j-k|\over
\sin{\pi\over 2N}(j+k)}.
\ee

To proceed we first perform the sum over k,
$$
\sum_{k=0}^{N-1}\sin {\pi ki'\over N}\sin {\pi kj'\over N}
\cos{\pi k(i-\2)\over N}\cos{\pi k(j-\2)\over N}=
\frac{1}{8}\sum_{\alpha\in \Delta_+}\sum_{k=0}^{N-1}\cos(\frac{\pi k\alpha}{N})
-
\frac{1}{8}\sum_{\alpha\in \Delta_-}\sum_{k=0}^{N-1}\cos(\frac{\pi k\alpha}{N})
$$
where $$\Delta_+=\{i'-j'\pm(i-j), i'-j'\pm(i+j-1)\},\qquad
\Delta_-=\{i'+j'\pm(i-j), i'-j'\pm(i+j-1)\}.$$
Now
$$\sum_{k=0}\sp{N-1}\cos(\frac{\pi k\alpha}{N})=
\cases{
\2[1-\cos\pi\alpha] &$\alpha\ne0, 2N$,\cr
N& $\alpha=0, 2N$.\cr }
$$
If every $\alpha\in\Delta_\pm$ is distinct from $0$ or $2N$, there is cancelling
between the terms in the sums. We find (taking account of parity) that
$$
\sum_{k=0}^{N-1}\sin {\pi ki'\over N}\sin {\pi kj'\over N}
\cos{\pi k(i-\2)\over N}\cos{\pi k(j-\2)\over N}=
\frac{N}{8}\left(
\sum_{\alpha\in \Delta_+} \delta_{\alpha,0}+\delta_{\alpha,2N}
-\sum_{\alpha\in \Delta_-}\delta_{\alpha,0}+\delta_{\alpha,2N}\right) .
$$
Without loss of generality we may assume $i'>j'$. Then
\be
\sum_{i,j=1}\sp{N}\sum_{\alpha\in \Delta_+}(\delta_{\alpha,0}+
\delta_{\alpha,2N}) {\tilde T}_{ij}^{\rm pert}
=
\sum_{i=1}\sp{N-(i'-j')}{\tilde T}_{i\, i+(i'-j')}\sp{\rm pert}+
\sum_{i=i'-j'+1}\sp{N}{\tilde T}_{i\, i-(i'-j')}\sp{\rm pert} \\
 +\sum_{i=N-(i'-j')+1}\sp{N}{\tilde T}_{i\, 2N-i-(i'-j')+1}\sp{\rm pert}+
\sum_{i=1}\sp{i'-j'}{\tilde T}_{i\,i'-j'+1-i}\sp{\rm pert}
\label{posroots}
\ee
In the limit $a_l\to 2\cos{\pi (l-\2)\over N}$ we note that
\be
{\tilde T}_{ij}\sp{\rm pert}={\tilde T}_{ij+2N}\sp{\rm pert}=
{\tilde T}_{i1-j}\sp{\rm pert}={\tilde T}_{i 2N+1-j},
\label{inds}
\ee
and for $i\ne j$
\be
{\tilde T}_{ij}\sp{\rm pert}=
\frac{1}{2i\pi}\log4\left(
\cos{\pi (i-\2)\over N}-\cos{\pi (j-\2)\over N}\right)\sp2\\
=\frac{1}{2i\pi}\log\left(
16\sin\sp2{\pi (i-j)\over 2N}\sin\sp2{\pi (i+j-1)\over 2N}\right)
\ee

To proceed we now distinguish two cases depending upon whether
$i'-j'$ is even or odd. Let us take the even case first.
Using (\ref{inds})
we may combine the first and third terms of the right hand side,
of (\ref{posroots})
and the second and last terms to give
\be
\label{evenc}
\sum_{i,j=1}\sp{N}\sum_{\alpha\in \Delta_+}(\delta_{\alpha,0}+
\delta_{\alpha,2N}) {\tilde T}_{ij}^{\rm pert}
=\sum_{i=1}\sp{N}{\tilde T}_{i\, i+(i'-j')}\sp{\rm pert}+
\sum_{i=1}\sp{N}{\tilde T}_{i\, i-(i'-j')}\sp{\rm pert}.
\ee
The sum over $ \Delta_-$ similarly simplifies, though we note that
here there are now distinct cases to be considered (depending on whether
$i'+j'<N$ or not). We obtain
\be
\label{evenab}\sum_{i,j=1}\sp{N}\sum_{\alpha\in \Delta_-}(\delta_{\alpha,0}+
\delta_{\alpha,2N}) {\tilde T}_{ij}^{\rm pert}
=\sum_{i=1}\sp{N}{\tilde T}_{i\, i+(i'+j')}\sp{\rm pert}+
\sum_{i=1}\sp{N}{\tilde T}_{i\, i-(i'+j')}\sp{\rm pert}
\ee
Combining our results shows (for $i'\ne j'$)
\be
S = \frac{N}{8} \sum_{i=1}\sp{N}\left(
{\tilde T}_{i\, i+(i'-j')}\sp{\rm pert}+
{\tilde T}_{i\, i-(i'-j')}\sp{\rm pert}-
{\tilde T}_{i\, i+(i'+j')}\sp{\rm pert}-
{\tilde T}_{i\, i-(i'+j')}\sp{\rm pert}\right) =
\\
=\frac{N}{16i\pi}\log\prod_{i=1}\sp{N}\left(
\frac{\sin\sp4{\pi (i'-j')\over 2N}\sin\sp2{\pi (2i+i'-j'-1)\over 2N}
\sin\sp2{\pi (2i-i'+j'-1)\over 2N} }
{\sin\sp4{\pi (i'+j')\over 2N}\sin\sp2{\pi (2i+i'+j'-1)\over 2N}
\sin\sp2{\pi (2i-i'-j'-1)\over 2N} }
\right) = \\
=  \frac{N^2}{4} T^D_{i'j'}
\ee
thus proving the conjecture for this case. Observe
that no terms in this product vanish with our assumption of $i'-j'$ being
even.

The remaining case follows in an analogous fashion, the difference now
being that $T_{ii}$ terms can appear when $i'-j'$ is odd. Set
$\delta=i'-j'$ and $\bar\delta=i'+j'$. Let $\bar\delta<N$
(a similar argument holding for $\bar\delta>N$).
With $\delta$ odd (\ref{evenc}) takes the form
\be
\label{oddc}
\sum_{i,j=1}\sp{N}\sum_{\alpha\in \Delta_+}(\delta_{\alpha,0}+
\delta_{\alpha,2N}) {\tilde T}_{ij}^{\rm pert}
=\sum_{i=1}\sp{N\prime}{\tilde T}_{i\, i+\delta}\sp{\rm pert}+
\sum_{i=1}\sp{N\prime}{\tilde T}_{i\, i-\delta}\sp{\rm pert}
+{\tilde T}_{ \frac{\delta+1}{2},\frac{\delta+1}{2}}\sp{\rm pert}+
{\tilde T}_{N-\frac{\delta-1}{2},N-\frac{\delta-1}{2}}\sp{\rm pert},
\ee
while (\ref{evenab}) becomes
\be
\sum_{i,j=1}\sp{N}\sum_{\alpha\in \Delta_-}(\delta_{\alpha,0}+
\delta_{\alpha,2N}) {\tilde T}_{ij}^{\rm pert}
=\sum_{i=1}\sp{N}{\tilde T}_{i\, i+\bar\delta}\sp{\rm pert}+
\sum_{i=1}\sp{N}{\tilde T}_{i\, i-\bar\delta}\sp{\rm pert}
+{\tilde T}_{ \frac{\bar\delta+1}{2},\frac{\bar\delta+1}{2}}\sp{\rm pert}+
{\tilde T}_{N-\frac{\bar\delta-1}{2},N-\frac{\bar\delta-1}{2}}\sp{\rm pert}.
\ee
Combining these two expressions yields
$$
S=\frac{N(N-1)}{4}T^D_{i'j'}+
\frac{N}{8}\left(
{\tilde T}_{ \frac{\delta+1}{2},\frac{\delta+1}{2}}\sp{\rm pert}+
{\tilde T}_{N-\frac{\delta-1}{2},N-\frac{\delta-1}{2}}\sp{\rm pert}-
{\tilde T}_{ \frac{\bar\delta+1}{2},\frac{\bar\delta+1}{2}}\sp{\rm pert}
-{\tilde T}_{N-\frac{\bar\delta-1}{2},N-\frac{\bar\delta-1}{2}}\sp{\rm pert}
\right).
$$
Finally the last four terms in brackets may be simplified (with
most terms cancelling) leaving $2 T^D_{i'j'}$. Again we arrive at
$$
S=\frac{N^2}{4}T^D_{i'j'}
$$
proving the conjecture.

\end{document}